\documentclass[journal=jacsat,manuscript=article]{achemso}
\usepackage[version=3]{mhchem}
\usepackage{balance}
\usepackage{mathptmx}
\usepackage{sectsty}
\usepackage{graphicx}
\usepackage{lastpage}
\usepackage{float}
\usepackage{fancyhdr}
\usepackage{fnpos}
\usepackage[english]{babel}
\usepackage{pdfpages}

\usepackage{array}
\usepackage{charter}
\usepackage[T1]{fontenc}
\usepackage{setspace}
\usepackage[compact]{titlesec}
\usepackage{hyperref}

\usepackage{epstopdf}
\usepackage{amssymb}

\title{Softness Matters: Effects of Compression on the Behavior of Adsorbed Microgels at Interfaces}

\author{Yuri~Gerelli}
\affiliation{Italian National Research Council - Institute for Complex Systems (CNR-ISC) and Department of Physics, Sapienza University of Rome, Piazzale Aldo Moro 5, 00185 Rome, Italy}
\email{yuri.gerelli@cnr.it}
\author{Fabrizio Camerin}
\affiliation{Division of Physical Chemistry, Lund University, P.O. Box 124, SE-22100 Lund, Sweden}
\email{fabrizio.camerin@gmail.com}
\author{Steffen~Bochenek}
\affiliation{Institute of Physical Chemistry, RWTH Aachen University, Landoltweg 2, 52056 Aachen, Germany}
\author{Maximilian~M.~Schmidt}
\affiliation{Institute of Physical Chemistry, RWTH Aachen University, Landoltweg 2, 52056 Aachen, Germany}
\author{Armando Maestro}
\affiliation{Centro de F\'{i}sica de Materiales (CSIC, UPV/EHU) - Materials Physics Center MPC, Paseo Manuel de Lardizabal 5, E-20018 San Sebasti\'{a}n, Spain.}
\alsoaffiliation{IKERBASQUE-Basque Foundation for Science, Plaza Euskadi 5, Bilbao, 48009 Spain.}
\alsoaffiliation{Institut Laue-Langevin, 71 Avenue des Martyrs, 38042 Grenoble, France}
\author{Walter~Richtering}
\affiliation{Institute of Physical Chemistry, RWTH Aachen University, Landoltweg 2, 52056 Aachen, Germany}
\author{Emanuela~Zaccarelli}
\affiliation{Italian National Research Council - Institute for Complex Systems (CNR-ISC) and Department of Physics, Sapienza University of Rome, Piazzale Aldo Moro 5, 00185 Rome, Italy}
\author{Andrea~Scotti}
\affiliation{Department of Biomedical Science, Faculty of Health and Society, Malmö University, SE-205 06 Malmö, Sweden}
\alsoaffiliation{Biofilms - Research Center for Biointerfaces, Malmö University, SE-205 06 Malmö, Sweden}
\email{andrea.scotti@fkem1.lu.se}

\begin{document}
\maketitle

\section*{Abstract}
Deformable colloids and macromolecules adsorb at interfaces, as they decrease the interfacial energy between the two media. The deformability, or softness, of these particles plays a pivotal role in the properties of the interface. In this study, we employ a comprehensive \emph{in situ} approach, combining neutron reflectometry with molecular dynamics simulations, to thoroughly examine the profound influence of softness on the structure of microgel Langmuir monolayers under compression.
Lateral compression of both hard and soft microgel particle monolayers induces substantial structural alterations, leading to an amplified protrusion of the microgels into the aqueous phase. However, a critical distinction emerges: hard microgels are pushed away from the interface, in stark contrast to the soft ones, which remain steadfastly anchored to it. Concurrently, on the air-exposed side of the monolayer, lateral compression induces a flattening of the surface of the hard monolayer. This phenomenon is not observed for the soft particles as the monolayer is already extremely flat even in the absence of compression.
These findings significantly advance our understanding of the pivotal role of softness on both the equilibrium phase behavior of the monolayer and its effect when soft colloids are used as stabilizers of responsive interfaces and emulsions.

\section{Introduction}
Colloidal particles and macromolecules are known to reduce interfacial energies, allowing them to be confined at the interface of immiscible liquids \cite{Bin02}.
This property has been extensively utilized to stabilize emulsions, employing surfactants \cite{Dek23}, proteins \cite{Sin11}, and polymers \cite{Hob18}.
Polymer crosslinked particles, i.e., microgels, are also often used to realize Pickering-like emulsions \cite{Ram04, Pic07, Hos21}.
If the polymer used in the syntheses has a lower critical solution temperature (LCST), e.g.~poly(N-isopropylacrylamide), the emulsions can be broken on demand by changing the temperature above the LCST \cite{Ric12, Des11, Kwo19}.
Unlike hard colloids, microgels exhibit significant deformations upon adsorption \cite{Pin14, Sco22Rev}, and their protrusion in the two subphases is sensitive to their softness \cite{Boc22}.
Moreover, the diverse compressibility within their volume \cite{Sch21, Hou22} gives rise to intriguing phase behavior, including the formation of different hexagonal lattices \cite{Rey16}.
In this framework, mechanical and visco-elastic properties of air-water and oil-water microgel-covered interfaces have been investigated by means of interfacial rheology and surface pressure measurements to determine how microgel conformation, packing and film stability change in response to compression \cite{Bru10,Ake17,Tat23,Sch23, Kaw23, Rey2023}.

Softness is known to strongly affect the properties of microgels at interfaces, such as the formation of crystalline lattice, the contact angle they form and the final architecture they assume once adsorbed \cite{Sco22Rev}.
The softest pNIPAM-based microgels that can be synthesized by precipitation polymerization \cite{Zhang1999, Saunders2004, Nishizawa2019, Nishizawa2021} are the so-called ultra-low crosslinked (ULC) microgels \cite{McP93, Bac15, Bru19}.
They exhibit properties that lie between those of hard particles and linear polymers \cite{Sco19, Boc22} and for this reason they play a crucial role in emulsion stabilization, enabling the formation of stable emulsions with properties in between those stabilized with macromolecules and those stabilized by responsive hard microgels \cite{Pet23}. Indeed, very recently, it was demonstrated that the limited deformability of soft microgels is key for the formation of stimuli-responsive emulsions \cite{Rey2023}.
The study also indicated that changes in the in-plane dimension of the microgels above and below the volume phase transition temperature (VPTT) cannot be responsible for emulsion destabilization.
Therefore, changes in the vertical structure of the microgel monolayers are pivotal for droplet stability and resistance against flocculation and coalescence interfacial behavior.
However, the experimental technique utilized by the authors did not prove the vertical distribution of the microgels under external stimuli since the observations were based on cryo-scanning electron microscopy (cryo-SEM) \cite{Rey2023}.

Recent interfacial measurements of microgels highlighted that they also share common properties with proteins and antibodies confined at air-water and oil-water interfaces \cite{Nus22}.
For instance, antibody monolayers show similar values of the surface elastic modulus compared to hard microgels \cite{Tei20} and, in both cases, the elastic properties of the monolayer are related to particle deformation and interaction upon adsorption suggesting similarities in the microscopic structures of microgels and bio-molecules at interfaces\cite{Woo23}.

Hard and soft colloids at the interface are also often used as model systems to study fundamental problems, such as those related to two-dimensional crystallisation.
For example, hard particles at the air-water interface have been used to investigate the melting and self-diffusion, in two dimensions, of neutral and charged hard spheres \cite{Zah99, Zah00, Kel17}.
Similarly, soft colloids have been instrumental in exploring the polymer-to-particle duality \cite{Sco19} and understanding the impact of particle softness and deformability on the phase behaviour of soft spheres in two dimensions at rest \cite{Rey16,Des14,Sch17,Sco22Rev} and under flow \cite{Sch23}.

To better understand both the monolayer interfacial phase behavior and the stability of the microgel-covered interface, fundamental for emulsion stability, one must consider not only the in-plane structure of the microgel but also how the out-of-plane morphology of these particles responds in different liquid (water and oil) and gas (air) phases.

Traditional techniques for characterizing the out-of-plane profile of microgels at interfaces rely on \emph{ex situ} analysis, involving monolayer deposition on solid substrates followed by methods like atomic force microscopy (AFM) \cite{Sch10, Cor17}, and freeze-fracture shadow-casting cryogenic scanning electron microscopy (FreSCa cryo-SEM) \cite{Gei14, Gei14b}.
However, these approaches lack the capability to study the \emph{in situ} properties of the monolayer, making it challenging to investigate dynamic responses to external conditions like compression forces or temperature changes.
Furthermore, the transfer of a microgel monolayer from a fluid interface to a solid substrate might introduce artifacts, raising concerns about the reliability of observed solid-to-solid isostructural phase transitions \cite{Kuk23, Kaw23}.

Among surface-sensitive techniques, neutron reflectometry (NR) can be used to determine the out-of-plane profile of the microgels orthogonal to the interface and to assess their volume fraction \emph{in situ} \cite{Zie16,Boc22}.
In this work, by combining NR, surface pressure measurements, and computer simulations, we comprehensively describe the out-of-plane behavior of pNIPAM-based microgel monolayers under compression at the air-water interface. Microgel particles spread at different fluid interfaces, such as air-water or oil-water for instance, are characterised by a similar behaviour as determined in the absence of lateral compression \cite{Rey2020, Boc21}. Switching from air to oil as the hydrophobic phase results only in a slightly greater protrusion of the polymer chains in the hydrophobic phase.  Hence, we anticipate our findings obtained in the presence of lateral compression to be applicable beyond the air-water interface investigated here, extending to other fluid interfaces.

To probe the effect of softness, pNIPAM microgels crosslinked with 5~mol\% \textit{N,N'}-methyl\-ene\-bis\-acryl\-am\-ide~(BIS) \cite{Boc22} and ultra-low crosslinked ones \cite{Boc22} are used. The latter are obtained without the addition of any crosslinker and the polymeric network forms due to the hydrogen atom abstraction at the tertiary carbon atom of the isopropyl group \cite{Bru19}. In accordance with the terminology from our recent works, we name the microgels prepared with ultra-low cross-linker content as ‘soft’ and those with high cross-linker content, having a bulk modulus two orders of magnitude larger than the ultra-low crosslinked ones \cite{Hou22}, as ‘hard’. In solution, these hard microgel particles feature a core-corona internal structure \cite{Sti04FF}, while the soft ones are characterised by a more homogeneous polymer network \cite{Bru19}.

NR experiments are performed on a null-reflecting interface using partially deuterated polymers to highlight the structural changes taking place at the nano-scale in the microgel Langmuir monolayer upon compression.
Furthermore, computer simulations are exploited to evaluate the structural changes taking place upon lateral compression for \emph{in silico} microgels synthesized at different cross-link contents \cite{Gna17}.

Our results show that despite similar surface elasticity of hard and soft monolayers, upon compression hard microgels are pushed further away from the interface, with a significant decrease in their protrusion in air and the formation of a thicker polymer-dense phase in water.
In contrast, soft microgels remain anchored at the interface: an increase in the density of the microgels in water is observed but at the same time, their protrusion in air is not affected by compression and remains constant and limited to few nanometers.
Furthermore, the the polymer volume fraction of pNIPAM sitting onto the interface is virtually the same for both the hard and soft microgels.

\section{Methods}

\noindent\textbf{Synthesis.} Deuterated 5~mol\% crosslinked and ultra-low crosslinked pNIPAM-based microgels measured here are the same as those investigated in our previous study at the water interface \cite{Boc22}.

Precipitation polymerization was used to synthesize deuterated regular 5 mol\% crosslinked microgels.
1.5072~g of a NIPAM monomer with 7 hydrogen atoms substituted by deuterium (D7-pNIPAM, [C$_6$D$_7$H$_4$NO]$_n$) was added to 0.1021~g of \textit{N,N'}-methyl\-ene\-bis\-acryl\-am\-ide~(BIS, crosslinker agent), and 20.2~mg of sodium do\-de\-cyl sulfate~(SDS) and dissolved in 83~mL of filtered double-distilled water.
The solution was heated to 60~$^\circ$C under constant stirring while purged with nitrogen for 1~h.
The reaction was started by rapid addition of the initiator, a degassed solution of 37.1~mg of potassium per\-oxy\-di\-sul\-fate~(KPS) in 5~mL water.
The reaction continued for 4~h under constant stirring before temperature was lowered to room temperature.

Deuterated ultra-low crosslinked (ULC) microgels were obtained without the addition of any crosslinker agent from free radical precipitation polymerization of D3-NIPAM (C$_6$D$_3$H$_8$NO)~\cite{Bru19}.
Briefly, the monomer solution consisted of 70 mmol/L of D3-NIPAM and 1.2 mmol/L SDS in water.
The reaction solution was purged with nitrogen under stirring at 100 rpm and heated to 70\,$^\circ$C.
At the same time, a solution of KPS~(1.6 mmol/L in the reaction solution) was degassed.
To initiate the reaction, the KPS solution was then transferred into the monomer solution.
The reaction proceeded for 4 hours under constant stirring at 70\,$^\circ$C before being stopped by decreasing the temperature to room temperature.
The use of a monomer with only 3 atoms of deuterium is needed since in the monomer with 7 deuterium atoms the isopropyl group of NIPAM is deuterated and this suppresses the hydrogen atom abstraction responsible for the formation of the polymeric network when no crosslinker is used during the synthesis \cite{Bru19}.

Both microgels were purified by threefold centrifugation and lyophilization was applied for storage.
The choice of using deuterated microgels was made to increase the signal originating from the particles at the air-water interface during NR experiments.

\noindent\textbf{Monolayer preparation and compression.} All measurements were performed at the air-liquid interface using a temperature-controlled PTFE Langmuir trough (Kibron, Finland) as sample environment.
Surface pressure $\pi$ was measured and monitored continuously using paper Wilhelmy plates.
The trough was equipped with a single PTFE movable barrier that slides, in contact with the top of the liquid phase, parallel to the walls of the trough.
By its movement, the area onto which monolayers were deposited could be varied from 160~cm$^2$ to 70~cm$^2$, being the smaller value defined by instrumental constraints such as the footprint of the neutron beam and the presence of the pressure sensor.
In all experiments, surface compression was unidirectional, aligned with the movement of the barrier.
The rate of compression was consistently set at $\simeq 6.5$ cm$^{2}$min$^{-1}$ for all samples, ensuring a gradual and uniform compression process.
This slow compression rate guarantees that the samples were probed under equilibrium conditions \cite{Sco19, Boc21, Boc22}.
Before and between measurements on different samples, the trough was cleaned and a fresh interface created.
The liquid phase consisted of a mixture of MilliQ-grade H$_2$O and D$_2$O (91.9:8.1, v/v), known as air contrast matched water (ACMW), which has a scattering length of zero (equal to that of air) and provides no contribution to the specular reflectivity \cite{Ca18}.
Using this approach, reflectivity originated only by the microgel particles localised at the interface.
Microgels dissolved at a concentration of 1~mg~mL$^{-1}$ in deuterated chloroform were spread drop-wise at the air-water interface with a Hamilton syringe.
The Langmuir monolayer was then compressed to the desired surface pressure and reflectivity was measured.
The measured surface pressure ranged from 1 to 30~mN~m$^{-1}$, but the entire interval could not be probed by a single compression.
Therefore, different initial amounts of the microgel solutions were spread onto fresh interfaces to reach all measurement points.
The temperature was set to T~=~20~$\pm$~0.5~$^\circ$C.
Quasi-static compression isotherms (Fig.~S1, ESI\dag) were measured to probe the mechanical response of the monolayers.
From the $\pi-A$  curves (being $A$ the area normalised by the amount of microgel), the compression elastic modulus of the surface, $\varepsilon$, was calculated as \cite{Dav63,Cicuta2005}

\begin{equation}
\varepsilon = - A \frac{d \pi}{d A}
\label{eq:Cs}
\end{equation}

\noindent\textbf{Neutron reflectometry.} NR measurements have been performed using the time-of-flight Fluid Interfaces Grazing Angles Reflectometer (FIGARO) at the Institut Laue-Langevin (ILL) in Grenoble, France \cite{Ca11}.
Briefly, reflectivity $R$, i.e., the ratio between the number of reflected and incident neutrons, is measured as a function of the exchanged wave-vector $Q$ in the direction perpendicular to the reference interface.
In first approximation, R is proportional to the Fourier transform of the first derivative of the scattering length density (SLD) with respect to $z$, which is the direction perpendicular to the horizontal plane defined by the air-water interface (located at $z=0$).

Since the SLD depends on the type of nuclei present in the sample, from the analysis of NR data, it is possible to determine the sample structure and the volume fraction profiles, $\phi(z)$, of its components.

The $Q$ range of interest was covered using two configurations with the incoming beam wavelengths $\lambda$ between 2 and 20~\AA~at two different angles of incidence, namely $\theta$: 0.615$^\circ$ and 3.766$^\circ$.
The $\frac{\Delta Q}{Q}$ resolution was set to 7\% for all the measurements and the footprint of the neutron beam at the sample position was kept constant at 1 $\times$ 4 cm$^2$.
Reflected neutrons were collected on a bi-dimensional $^3$He detector and converted to reflectivity curves by using the COSMOS routine provided by the ILL \cite{Gut18}.

\noindent\textbf{NR data modeling.} The main objective of NR data analysis was to accurately represent the projection of the microgel volume fraction profile ($\phi_{\mu g}$) along the vertical axis ($z$).
It is well-known that individual microgel particles in the swollen state in solution can be described using the fuzzy sphere form factor \cite{Sti04FF} characterized by a Gaussian-like decay of the radial polymer distribution in the corona region.
On the contrary, microgel particles do not maintain a spherical shape when adsorbed at an interface, either in dilute or crowded conditions \cite{Gei12,Pin14,Sco22Rev} and their interactions with the neighbouring particles in the monolayer can induce further deformations.
This, together with the presence of internal regions (core and corona) characterised by different polymer concentration, and therefore softness \cite{Sch21}, makes it challenging to develop an analytical model able to accurately describe the particle morphology at an interface.
Moreover, given the large size of swollen microgel particles with respect to the typical distances probed by neutrons in NR experiments, the use of the common representation of the monolayer in terms of a finite number of slabs \cite{Ger20} might not be sufficient to quantitatively describe the changes taking place in the microgel volume fraction profile (VFP) under compression.

In this work, these limitations are circumvented by using a phenomenological description of the projection of the microgel volume fraction profile $\phi_{\mu g}$ orthogonal to the interface.
Based on previous experimental and computational evidences \cite{Boc22}, a microgel monolayer is characterised by a denser region at the interface ($z=0$ nm), modeled in our approach by a Gaussian peak (Eq.~S1, ESI\dag), and by two regions, in water and in air, with lower polymer density that also gradually decreases to zero moving away from the interface.
These two regions, called protrusions in the manuscript, are modeled using error functions (Eq.~S2 and Eq.~S3, ESI\dag).
As detailed in the ESI\dag, the width, the position along $z$ and the amplitude of the error functions, as well as the position along $z$, width and amplitude of the Gaussian peak are the model parameters optimised during data analysis. The VFPs of air and water phases, namely $\phi_{air}(z)$ and $\phi_{w}(z)$ (Eqs.~S4 and~S5, ESI\dag), are described using error functions centered at $z=0$ with a fixed width of 0.3 nm, matching the expected surface roughness due to capillary waves in a free water surface at ambient pressure and temperature \cite{Bra85}.
Generic VFPs for air, water and microgel layer are shown in Fig.~S2 in the ESI\dag.
The total SLD profile $SLD(z)$ was computed by multiplying each VFP contribution by the respective material SLD value as
\begin{eqnarray}     \label{eq:SLD}\nonumber
    SLD(z) &=& SLD_{air}\phi_{air}(z) + SLD_{w}\phi_{w}(z) \\
     &+& SLD_{\mu g}\phi_{\mu g}(z).
\end{eqnarray}
In the present study, by using ACMW as liquid phase ($SLD_{air}=SLD_{w} = 0$), Eq.~\ref{eq:SLD} was simplified to $SLD(z)=SLD_{\mu g}\phi_{\mu g}(z)$, where $SLD_{\mu g}$ is the SLD value for the microgel determined experimentally by means of small-angle neutron scattering, namely $3.03\times 10^{-6}$ \AA$^{-2}$ for D3-pNIPAM \cite{Sco20ULC} and $5.39\times 10^{-6}$ \AA$^{-2}$ for D7-pNIPAM \cite{Sco21VF}, respectively.

As detailed in the literature\cite{Arm22}, reflectivity curves $R$ can be computed from an SLD profile by dividing the latter into a large number of finite-size slabs with zero interfacial roughness, which are then used as input for algorithms based on the Parratt's formalism \cite{Par54} or on the Abeles matrix method \cite{Abe50}.
In the present case, calculation of SLD profiles and of reflectivity curves was performed using algorithms, based on the Parratt's formalism, present in the Aurore software application \cite{Ger16} and the computed reflectivity was then fitted to the experimental data using a least squares minimization approach.
During this step, the smearing effect due to the instrumental resolution as well as the presence of a flat background were taken into account. This iterative procedure led to the optimization of the parameters characterising the microgel VFP.

To enhance the reliability of the obtained parameters and to highlight the sensitivity of the proposed model to the different regions of the experimental data, the bootstrapping technique described in Ref. \cite{Ger16} was used.
One hundred bootstrap samples were generated by resampling the experimental data with replacement.
In this context, for every experimental point, a new reflectivity value was randomly selected within a range defined by the absolute experimental error.
For each bootstrap sample, the model reflectivity curves were optimized independently.
This procedure, despite being very intensive from a computational point of view, allowed to obtain a distribution of model curves representing 95\% confidence intervals (1.96 standard deviations for normal distributions).
Using all the models falling within this interval, the corresponding confidence intervals associated with the VFP were calculated.
An illustrative  example of the procedure outcome is presented in Fig.~S4$^{\dag}$.


\noindent\textbf{Modeling and interaction potentials.} \textit{In silico} microgels are designed to reproduce in a coarse-grained manner a standard pNIPAM network.
In the presence of BIS acting as crosslinking agent, we assemble the network by exploiting patchy particles with two and four patches, representing monomers and crosslinkers, respectively.
We focus on microgels with $c=5$\% of crosslinkers as in experiments, using a total number of particles $N \approx 5000$ and $N \approx 42000$.
Four patches particles also experience an additional design force to concentrate them more in the center of the network\cite{Nin19}.
The spherical shape of the microgel is obtained by applying a spherical confinement of radius $Z=25\sigma$, with $\sigma$ the unit of length in simulations and the size of each particle.
For a fast assembly, we make use of the \textsc{oxdna} simulation package~\cite{rovigatti2015comparison} on GPU processors.
As already reported in our previous work \cite{Boc22, Nab23}, for the ULC microgels, we use the same approach based on patchy particles, but the fraction of crosslinkers, amounting to self-crosslinks among pNIPAM monomers, is equal to $c=0.3$\%, as estimated from a comparison with experimental data in Ref.~\cite{Nab23}.
In this case, we use $N \approx 21000$ and a much lower network density, with the spherical confinement taking place in a sphere of radius $Z=55.5\sigma$.
More details on the assembly protocol and on the interaction potentials for patchy particles can be found in Refs.~\cite{gnan2017silico,Nin19,Boc22, Nab23}.

Subsequently, for preserving the topology of the assembled patchy particles, each link is substituted with a permanent bond as in the Kremer-Grest bead-spring model for polymers.
In this way, all beads interact \emph{via} the Weeks-Chandler-Anderson (WCA) potential
\begin{equation}
V_{\rm WCA}(r)=
\begin{cases}
4\epsilon\left[\left(\frac{\sigma}{r}\right)^{12}-\left(\frac{\sigma}{r}\right)^{6}\right] + \epsilon & \text{if $r \le 2^{\frac{1}{6}}\sigma$}\\
0 & \text{otherwise.}
\end{cases}
\end{equation}
\noindent with $\epsilon$ setting the energy scale and $r$ the distance between two particles. Connected beads  also interact \emph{via} the Finitely Extensible Nonlinear Elastic (FENE) potential,
\begin{equation}
V_{\rm FENE}(r)=-\epsilon k_FR_0^2\ln\left[1-\left(\frac{r}{R_0\sigma}\right)^2\right]     \text{ if $r < R_0\sigma$,}
\end{equation}

\noindent with $k_{\rm F}=15$ which determines the stiffness of the bond and $R_{\rm 0} = 1.5$ is the maximum bond distance.

To study \textit{in silico} microgels at an interface, we then make use of explicit solvent particles. These are treated as soft beads within the dissipative particle dynamics (DPD) framework~\cite{groot1997dissipative,camerin2018modelling}.
Therein, the total interaction force among beads is $\vec{F}_{\rm ij} = \vec{F}^C_{\rm ij} + \vec{F}^D_{\rm ij} + \vec{F}^R_{\rm ij}$, where:
\begin{eqnarray}
	\vec{F}^C_{ij}  &=&  a_{ij} w(r_{ij}) \hat{r}_{ij} \\
	\vec{F}^D_{ij}  &=&  -\gamma w^2(r_{ij}) (\vec{v}_{ij}\cdot\vec{r}_{ij}) \hat{r}_{ij} \\
	\vec{F}^R_{ij}  &=&  2\gamma\frac{k_B T}{m} w(r_{ij}) \frac{\theta}{\sqrt{\Delta t}} \hat{r}_{ij}
\end{eqnarray}
As from previous works~\cite{Cam19,Cam20}, we set $a_{\rm 11}=a_{\rm 22}=8.8$, $a_{\rm 12}=31.1$, for the interactions between fluid 1 and fluid 2.
Instead, for the monomer-solvent interactions, $a_{\rm m1}=4.5$ and $a_{\rm m2}=5.0$, making fluid 1, representing water, the preferred phase. The cut-off radius was always set to be $r_{\rm c}=1.9 \sigma$, the reduced solvent density $\rho_{\rm DPD}=4.5$.

\noindent\textbf{System setup and simulation details.} For the microgels with $c=5$\%, we create a monolayer of microgels with $N=5000$ monomers by disposing twelve of them in a rectangular simulation box with periodic boundary conditions.
We first let the system equilibrate at low densities in the presence of explicit solvent, so that the particles acquire the characteristic ``fried-egg" shape experimentally observed once they are adsorbed onto interfaces \cite{Rey2023, Gei14, Sch21, Sco22Rev}.
We then progressively reduce the size of the simulation box in the $x$ and $y$ directions, leaving unaltered the $z$ direction, perpendicular to the plane of the interface, and keeping constant the solvent density.

We also perform compression runs of single microgels of larger size ($N=42K$ beads for $c=5$\% and $N=21K$ beads for ULC ones) by imposing an external force of cylindrical symmetry, with $z$ as the main axis.
In this way, all the monomers experience a harmonic force $F(r)=-k(r-R)^2$ with $F(r)=0$ if $r>R$, where $r$ is the distance from the monomer to the center axis of the cylinder, $R$ is the equilibrium radius of the cylinder and $k=10$ is the intensity of the force.
Solvent beads are not subjected to $F(r)$.
The increased size of ULC microgels in simulations makes it unfeasible with present computational resources to run multiple microgels or larger system sizes for the single ones, due to the larger box size filled with solvent molecules.

Simulations are run in the NVT ensemble fixing the reduced temperature of the system $T^*=1$  \emph{via} the DPD thermostat, and they are performed with \textsc{lammps}~\cite{Pli95}.
In all cases, we record the microgel density profile $\rho(z)$, that is obtained by dividing the simulation box along the $z$ axis into three dimensional bins that are parallel to the interface. For the case of the microgel monolayer, we measure $\rho(z)$ for each microgel and then average it over all particles.

\noindent\textbf{Comparison between simulated systems and experiments.} A quantity that can be easily computed, allowing a direct comparison between simulated systems and those measured experimentally, is the generalized area fraction $\zeta_{2D}$ \cite{Sco19, Boc21, Sco22Rev, Sch23}:
\begin{equation}
\label{eq:GAF}
\zeta_{2D} = \frac{N_p \cdot A_{p}}{A_{\text{meas}}}
\end{equation}
where $N_p$ is the number of particles in the observed area $A_{\text{meas}}$, and $A_{p}$ is the interfacial area of an individual microgel particle in the dilute regime.
In the experiments, $A_{\text{meas}}$ was fixed to 4 cm$^2$, while $N_p$ increased upon compression, and the changes in $N_p$ are directly proportional to the changes in the area of the trough when compressed by the barrier movement.
In the simulations, $N_p$ was fixed, and $A_{\text{meas}}$ decreased upon compression.
In the simulations, the area of the individual microgel was determined, in the absence of compression, by the distribution of beads at $z=0$.
For the experiments, this value is usually determined ex-situ, i.e., by AFM on dry films transferred onto a solid substrate \cite{Boc21}.
Since the Langmuir trough experiments presented in the manuscript were performed in-situ, the simultaneous deposition of the monolayer and, consequently, the AFM measurements were not possible.
However, it has been reported that both ULC and hard microgel particles make contact at $\zeta_{2D}\approx 1$, and generally, this area fraction corresponds to the lift-off of the pressure-area isotherms \cite{Boc21, Sco22Rev, Sch23}.
This correspondence was therefore used, together with the changes in the $A_{meas}$ (of the simulation box and of the trough), to establish an equivalence between simulated and measured systems (see diamonds and stars in Fig.~S2, ESI\dag).
Simulations for hard microgels are the analogues of experiments performed at $\pi$ = 1, 19, 25, and 28 mN m$^{-1}$, while those performed on ULC correspond to experiments performed at $\pi$ = 1, 16, 25, 28, and 30 mN m$^{-1}$.

\section{Results}
\subsection{Surface pressure behaviour}
\begin{figure}[!ht]
    \center
        \includegraphics[width=.49\textwidth]{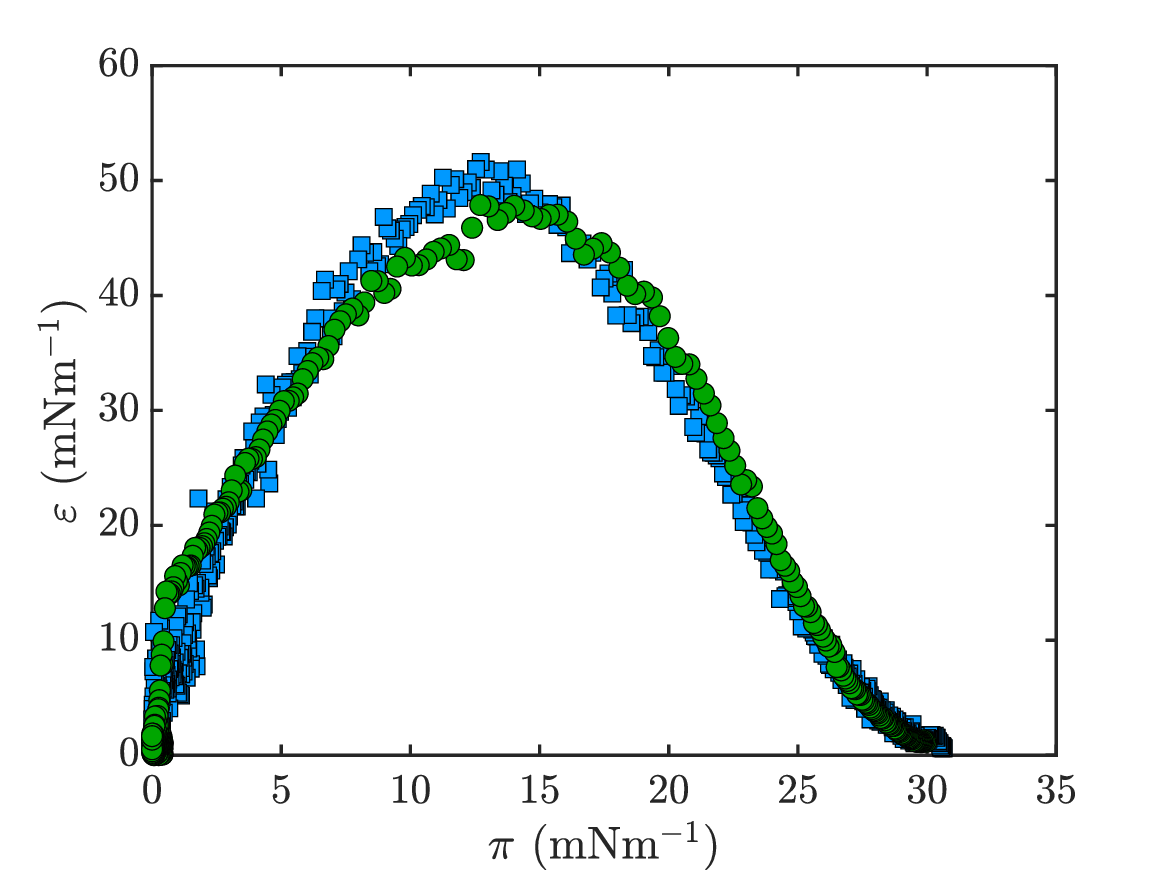}
        \caption{\footnotesize Compression elastic modulus $\varepsilon$ calculated using Eq. \ref{eq:Cs} as a function of the surface pressure $\pi$ for the hard (D7-pNIPAM, squares) and soft (D3-pNIPAM, circles) microgels.}
        \label{fig:Cs}
\end{figure}

The compression elastic modulus of the surface, $\varepsilon$, was calculated using Eq.~\ref{eq:Cs} from the compression isotherms shown in Fig.~S1$^{\dag}$.
Its dependence on the surface pressure $\pi$ is shown in Fig.~\ref{fig:Cs} for both hard and soft microgel monolayers.
During the compression, the elasticity of both monolayers first increases, up to a maximum of 49 $\pm$ 3 mN~m$^{-1}$ for surface pressure in the range 12 - 16 mN~m$^{-1}$, then it decreases down to zero as $\pi$ approaches 30 mN~m$^{-1}$.
The maximum elasticity occurs at slightly higher surface pressure for the soft microgel monolayer (circles), namely at 14.4 mN~m$^{-1}$ versus 13.0 mN~m$^{-1}$ for the hard particles (squares).

This behavior is similar to what was reported by Picard \emph{et al.} \cite{Pic17} and Pinaud \emph{et al.} \cite{Pin14} for pNIPAM monolayers at the air-water and at the oil-water interface respectively, using particles with cross-linker concentration varying from 1~mol\% \cite{Pic17} to 5~mol\% \cite{Pic17,Pin14}.
In both studies, the authors attributed the maximum surface elasticity to the flattened conformation of the microgel particles at the interface.
Despite the similar trend of $\varepsilon(\pi)$, the structure and arrangement of the hard and soft monolayers investigated in this study are different and may play a critical role in their overall behavior at the interface.

\subsection{Volume fraction profiles: neutron reflectometry}
\begin{figure}[!ht]
    \center
        \includegraphics[width=.49\textwidth]{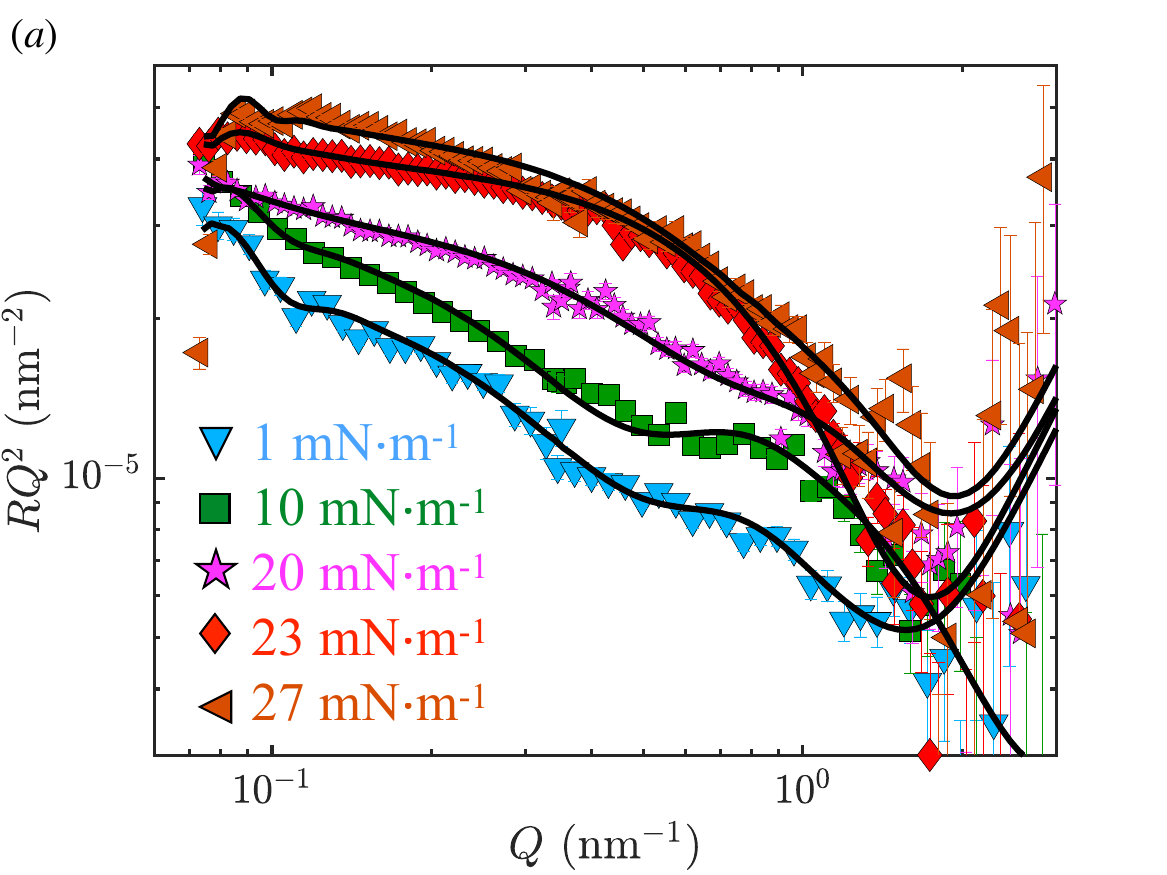}
        \includegraphics[width=.49\textwidth]{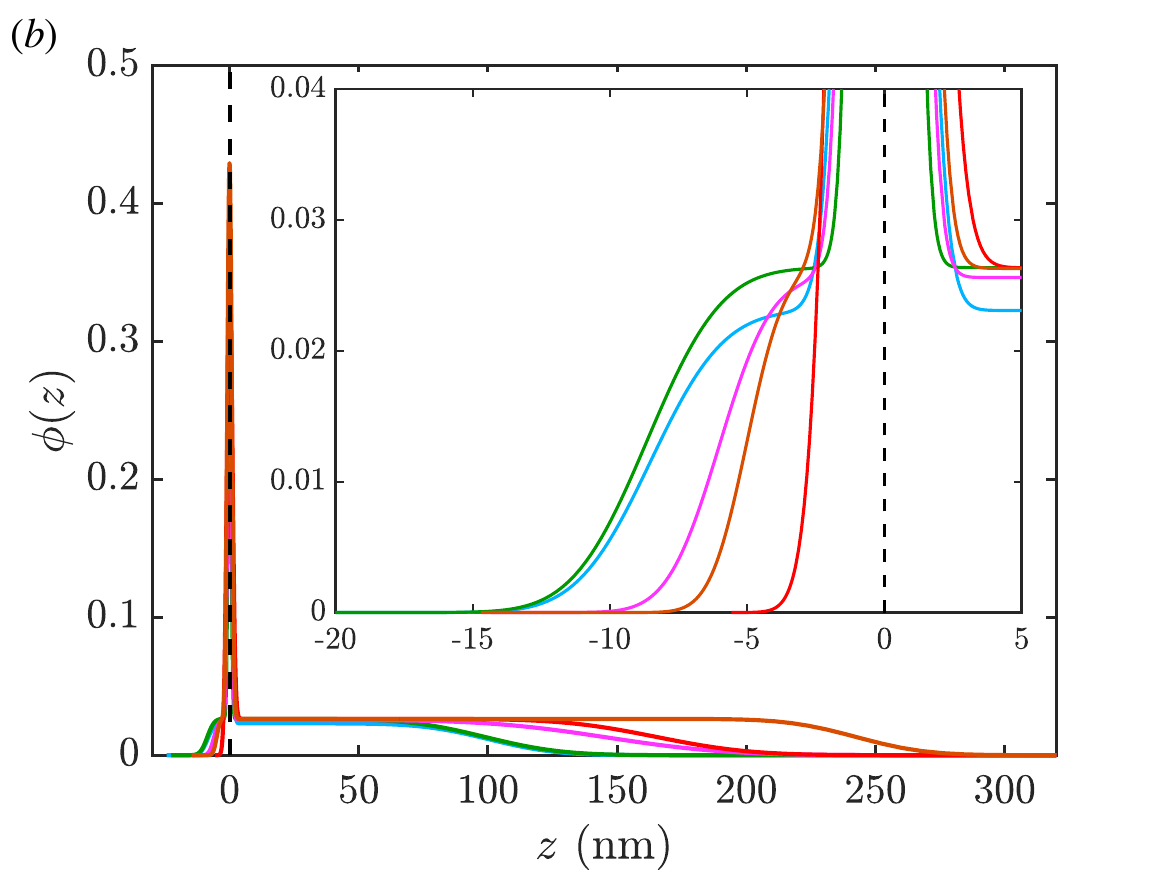}
    \caption{\footnotesize (a) Neutron reflectivity $R$ multiplied by $Q^2$ plotted as a function of the exchanged wave-vector $Q$ at different compressions of the hard microgel monolayer. The different symbols and colors correspond to different compression of the monolayer, as indicated in the legend. Solid lines represent model reflectivity curves originating from Eq.~\ref{eq:SLD}.
    All measurements were performed at T~=(20.0~$\pm$~0.5)~$^\circ$C.
    (b) Polymer volume fraction profiles as a function of the distance from the interface located at $z=0$ nm (dashed vertical line), corresponding to the model NR curves displayed in panel (a). Same colors are used in (a) and (b).}
        \label{fig:D7}
\end{figure}

Fig.~\ref{fig:D7}(a) shows the measured reflectivity, $R$, (symbols) for the hard microgel monolayer measured using ACMW as liquid phase and deuterated (D7-pNIPAM) microgel particles along with the model curves obtained from the data analysis (lines). Data are shown on a $RQ^2$ scale to highlight the oscillations present in the entire $Q$-range investigated.
The experimental data were collected upon compression of the monolayer at surface pressure values indicated in the figure legend.

At first glance, it is evident that reflectivity increases from the lowest (represented by down-side triangles) to the highest levels of applied compression (indicated by circles and left-side triangles).
This rise in reflectivity can be attributed to the increased polymer content within the area illuminated by the neutron beam, resulting from the lateral movement of the trough barriers.
Notably, in the very low-$Q$ regime, changes in reflectivity are of a similar order of magnitude as the changes in the illuminated area.
However, it is important to note that it is challenging to establish a precise proportionality between these values due to their dependence on the formation of a film. For the $Q$-range examined, this film cannot be considered extremely thin and leads to the presence of fringes in the reflectivity \cite{Cam16}.

In addition to changes in overall intensity, the shape of the experimental curves also changes in response to monolayer compression, indicating variations in the monolayer volume fraction profile $\phi_{\mu g}(z)$ and, consequently, in the sample scattering length density, SLD$(z)$, both defined along the direction $z$ orthogonal to the interface.
Indeed, the analysis of NR data, performed using the model described in the ESI\dag\ and summarised by Eq.~\ref{eq:SLD}, allowed us to determine $\phi_{\mu g}(z)$ uniquely (Fig.~\ref{fig:D7}(b)).
The chosen model effectively replicates the experimental data across the entire $Q$-range investigated and accurately captures fringes, common for thick films, visible at low-$Q$.  It is worth mentioning that the characteristic core-corona structure of microgel particles is preserved once they are spread at a fluid interface. However, due to effects such as deformation and interpenetration of polymer chains, as well as a limited difference in scattering length density, building a model capable of distinguishing these two regions within a microgel particle layer poses significant challenges.

The value of $\phi_{\mu g}(z=0)$ at the interface, given by the sum of the amplitude of the Gaussian peak function, $A_g$, and of the error functions, $A_d$, (see Eqs.~S1-S3 and schematic volume fraction profile in Fig.~S2(b), ESI\dag) increases from $0.24 \pm 0.02$ ($\pi = 1$~mN~m$^{-1}$) to $0.39 \pm 0.02$ ($\pi = 20$~mN~m$^{-1}$) as a function of compression.
This indicates the formation of a hydrated microgel monolayer and an increase in the number of microgel particles exposed to the neutron beam.
With further compression of the monolayer, $\phi_{\mu g}(0)$ slightly decreases to $0.36 \pm 0.02$ ($\pi = 1$~mN~m$^{-1}$).
Simultaneously, the interfacial peak broadens. Together, the two observations indicate the rearrangement of the polymer chains being pushed away from the surface. Finally, at the higher surface pressure investigated, the amount of material in the sample area increases ($\phi_{\mu g}(0)=0.43\pm0.02$) while the width of the peak remains constant.

\begin{figure}[!ht]
    \center
        \includegraphics[width=.5\textwidth]{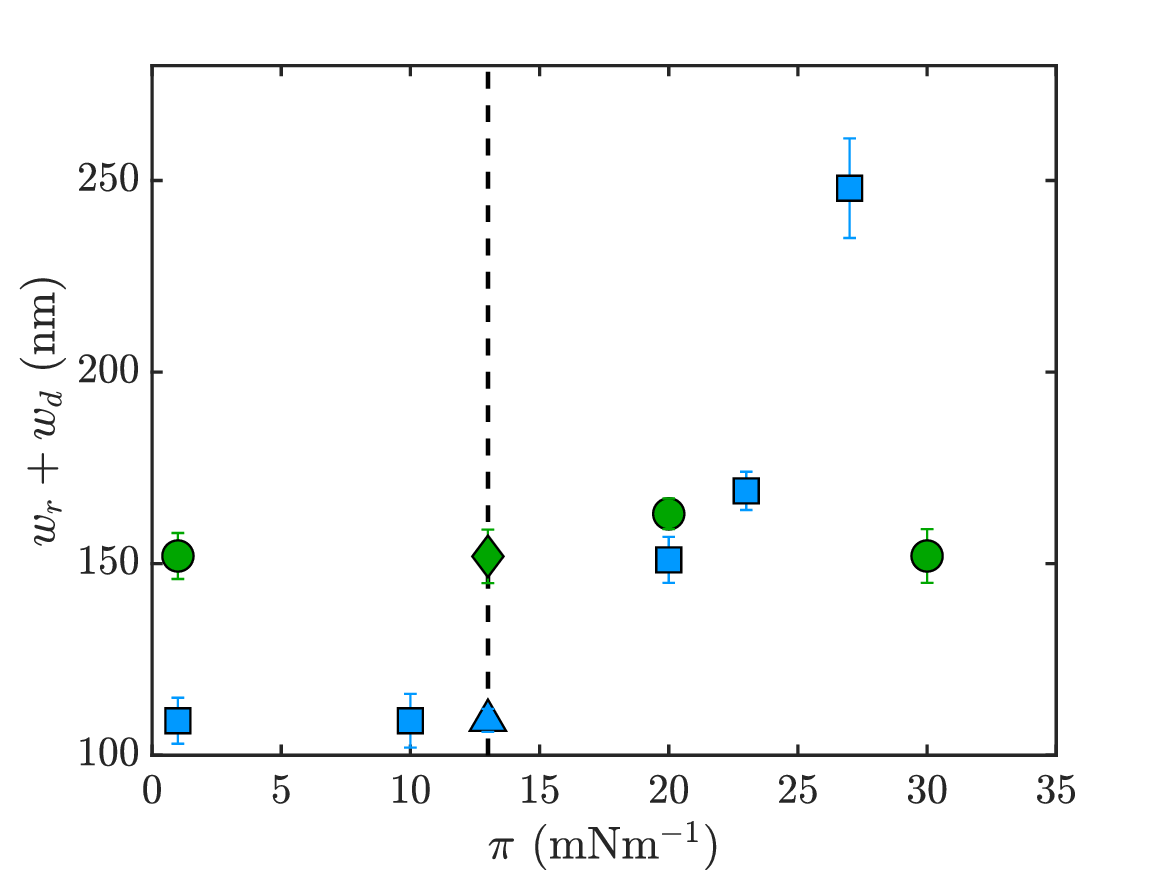}
    \caption{\footnotesize {Total extension of the VFP across the interface ($w_r+w_d$) for hard (squares) and soft (circles) microgels as a function of the surface pressure $\pi$.
    Values corresponding to a surface pressure of $\pi=13$ mN~m$^{-1}$ (triangle and diamond for the hard and soft microgels, respectively) were derived, using the model described in the present manuscript, from NR data published in \cite{Boc22}.}}
        \label{fig:wrwd}
\end{figure}

As described in the ESI\dag, the full extension of the microgel across the interface is given by $w_d+w_r$; it remains almost constant at $109 \pm 2$~nm for surface pressures in the range $1\le\pi\le 13$~mN~m$^{-1}$, while it continuously increases for larger compressions, reaching $240 \pm 10$ nm at $\pi = 27$~mN~m$^{-1}$ (squares and triangle in Fig.~\ref{fig:wrwd}).

Although the most significant changes in $\phi_{\mu g}(z)$ occur in the water phase, the model also accurately captures the differences in the protrusion of the microgel into the air phase (represented by $w_r$ in Fig.~S2(b), ESI\dag).
Notably, measurements on the microgel protrusion in hydrophobic phases are limited in the literature and are commonly conducted using FreSCa cryo-SEM \cite{Gei12, Gei14, Cam19, Sch17}, a technique that requires freezing and fracturing of the interface, actions that might not preserve the original monolayer properties \cite{Kuk23}. Only recently, the shape of microgel particles at a fluid oil-water interface was characterised by AFM \cite{Via22}.
In the inset of Fig.~\ref{fig:D7}(b), it can be observed that the microgels, upon initial interaction, protrude approximately $9\pm1$ nm into the air, a value comparable to what was previously reported by Bochenek et al.~\cite{Boc22}.
As compression is increased, the microgels are pushed away from the air, leading to a decrease in their protrusion to $5\pm1$ nm (solid triangles in Fig.~S3, ESI\dag).
It is important to note that, as illustrated in Fig.~S2$^{\dag}$, the protrusion is defined by the inflection point in the error function that describes the rise of the volume fraction profile.
When the stretching parameter $\alpha_r$ is large, the profile extends further into the air phase, as is evident in Fig.~\ref{fig:D7}(b) for the hard microgels at low compression.

As already mentioned and also reported in the literature \cite{Pet23}, the elastic response to compression of an ultra-low cross-linked microgel monolayer is virtually the same as that of a hard microgel monolayer in the pressure range investigated in the current work.
However, NR measurements indicate substantial differences in the organisation and deformation of soft microgel particles at the interface.

\begin{figure}[!ht]
    \center
        \includegraphics[width=.49\textwidth]{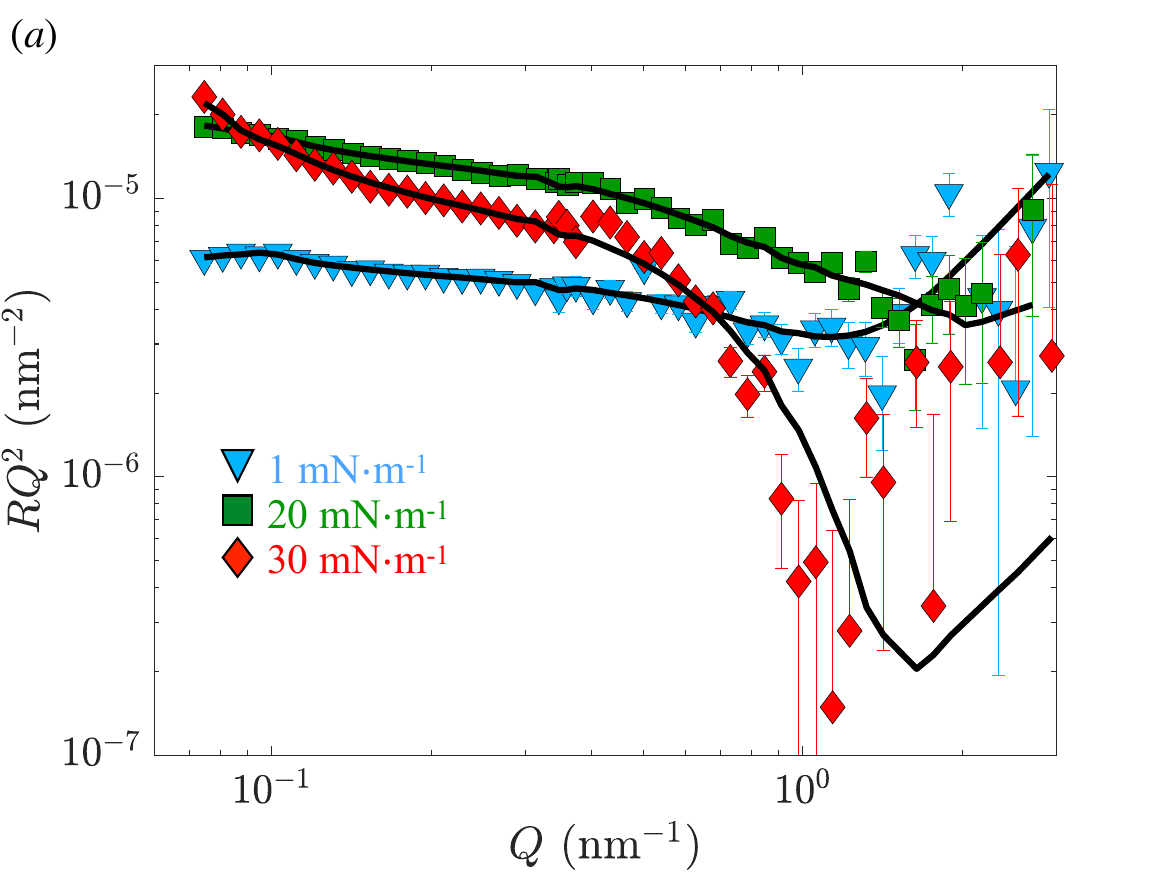}
        \includegraphics[width=.49\textwidth]{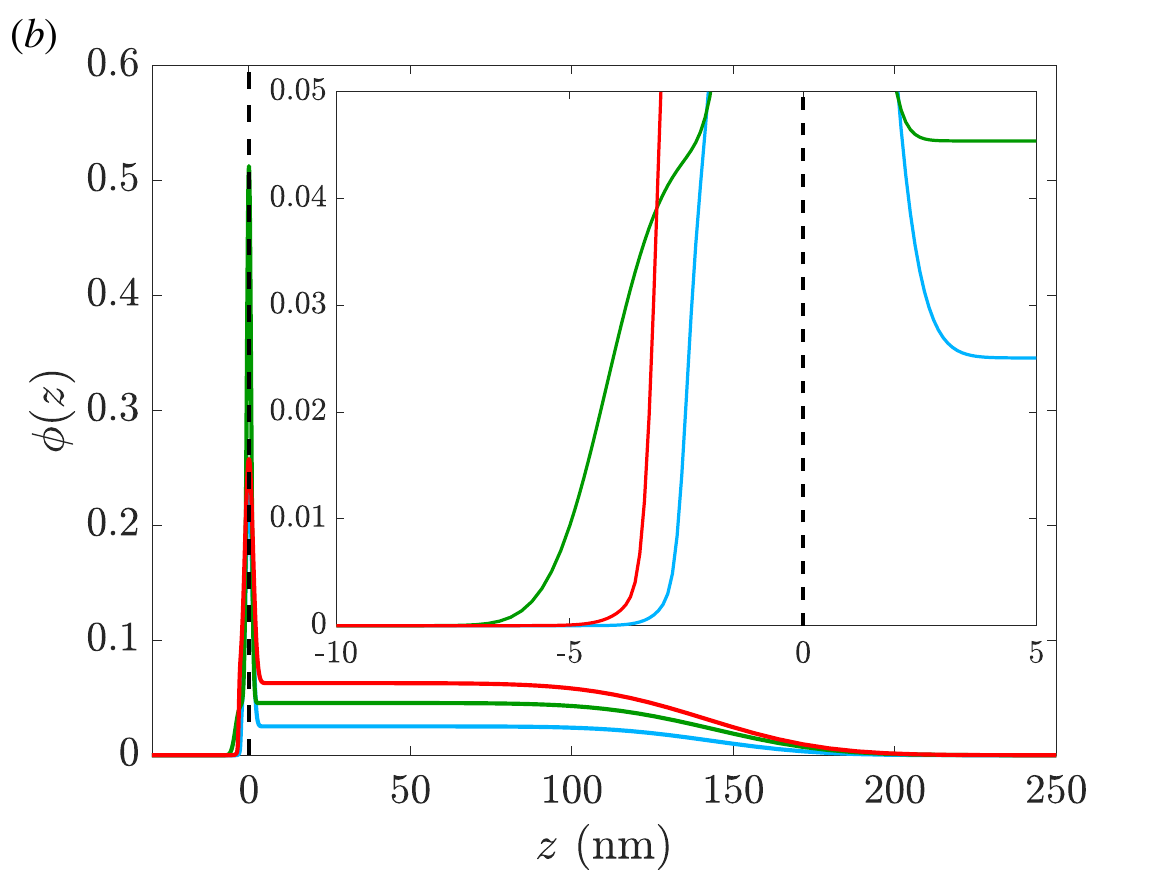}
        \caption{\footnotesize (a) Neutron reflectivity $R$ multiplied by $Q^2$ plotted as a function of the exchanged wave-vector $Q$ at different compressions of the soft microgel monolayer.
        The different symbols and colors correspond to different compression of the monolayer, as indicated in the legend. Solid lines represent model reflectivity curves originating from Eq.~\ref{eq:SLD}.
        All measurements were performed at T~=(20.0~$\pm$~0.5)~$^\circ$C.
        (b) Polymer volume fraction profiles as a function of the distance from the interface located at $z=0$ nm (dashed vertical line), corresponding to the model NR curves displayed in panel (a).
        Same colors are used in (a) and (b).}
        \label{fig:D3}
\end{figure}

The NR curves in Fig.~\ref{fig:D3}(a) show that the reflectivity measured for the soft microgel monolayer is generally lower than that measured for their hard counterpart at the same (or very similar) surface pressure values, leading to NR data characterised by a lower S/N ratio.
This difference is mainly due to the fact that ultra-soft microgels consist of less polymer \cite{Sco19} and that deuteration was performed only for three $^1$H atoms instead of seven. Since reflectivity is proportional to the square of the contrast, a significant amount of soft microgel particles at the interface would be necessary to achieve a reflectivity of the same order of magnitude as that measured for the hard microgel particles. Therefore, differences in the magnitude of the reflectivity alone cannot be reliably used to determine structural changes in the soft microgel monolayer. Instead, structural changes were quantified employing the same model used for the hard microgels. The theoretical reflectivity curves and the corresponding $\phi_{\mu g}(z)$ profiles obtained from the analysis are shown in Fig.~\ref{fig:D3}(a) (solid lines) and Fig.~\ref{fig:D3}(b), respectively.

At the onset of the interaction between microgel particles, denoted by the increase of the surface pressure from 0~mN~m$^{-1}$ to 1~mN~m$^{-1}$, $R$ is weak and almost featureless.
At larger compressions, the reflectivity increases and the shape of the experimental data indicates the formation of an interfacial monolayer.
This observation is reflected in the monolayer volume fraction profiles measured at different surface pressures.
In particular, the value attained by $\phi_{\mu g}(z=0)$ increases from 0.24 $\pm$ 0.05 for $\pi=1$~mN~m$^{-1}$ to 0.57 $\pm$ 0.05 for $\pi=13$~mN~m$^{-1}$ after which it decreases to its initial value with further compression.
The amount of polymer present in the interfacial region of the profile, i.e.~within $\pm$ 3$w_g$ from $z=0$, is however increasing continuously because of the increase of the width, $w_g$, of the Gaussian function.
Conversely, the amplitude of the volume fraction profiles in water and air, denoted by $A_{r}$ in Eqs.~S2\dag\ and~S3\dag, steadily increases from 0.02 $\pm$ 0.01 to 0.06 $\pm$ 0.01.
In the case of the hard microgel monolayer, this parameter did not show any large variation, remaining almost constant in the range 0.02 - 0.03 for all investigated pressures.
Furthermore, in the case of the soft microgel monolayer, the protrusion of particles into the air and water phases remains nearly constant, approximately 3.5 nm and 148 nm, respectively (green circles in Fig.~\ref{fig:wrwd}, and blue circles and red triangles in Fig.~\ref{fig:wrwd}).

\begin{figure*}[!ht]
    \center
        \includegraphics[width=0.95\textwidth]{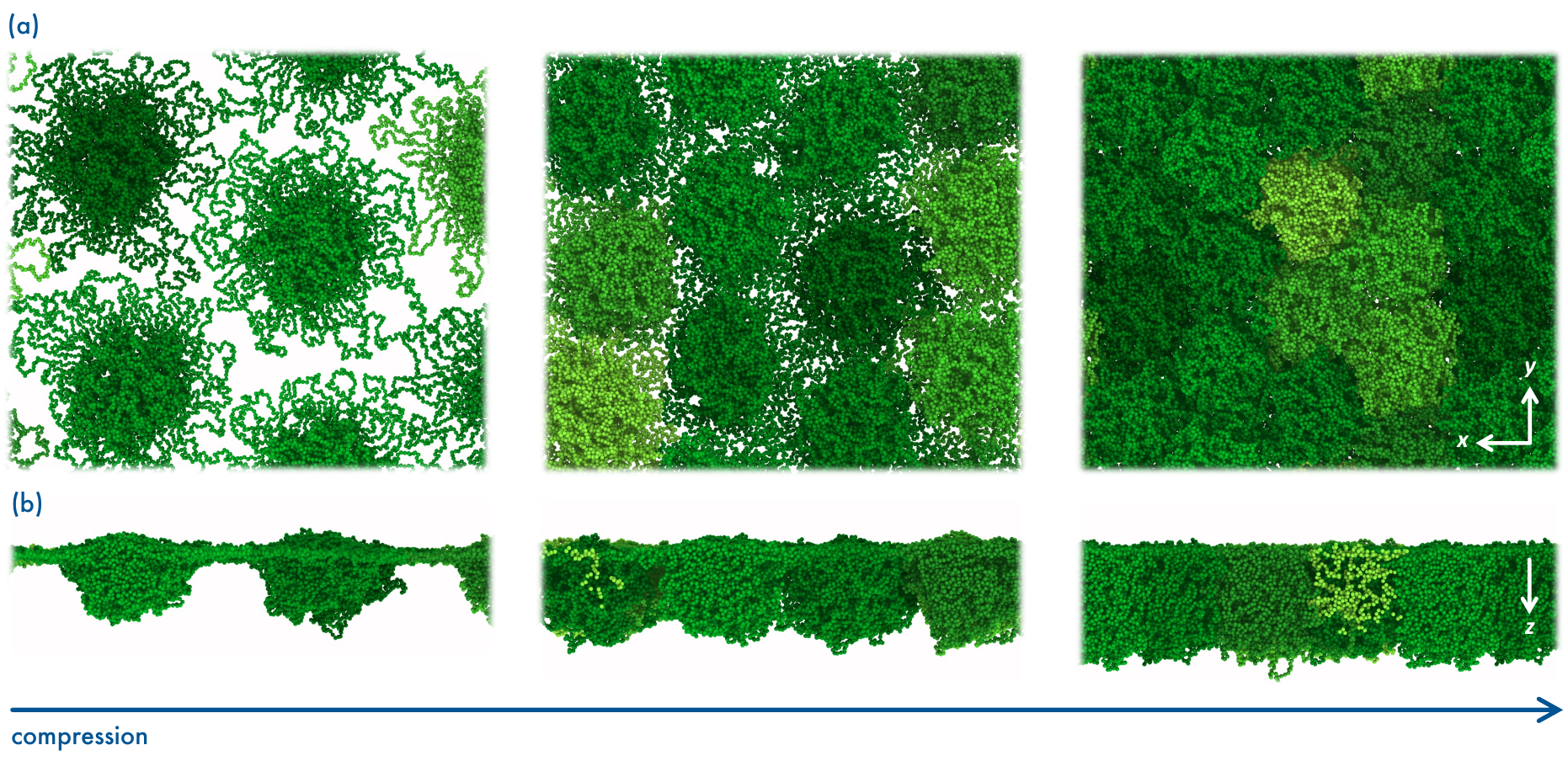}
        \caption{\footnotesize Representative simulations snapshots showing an ensemble of $5\%$ crosslinked microgels (hard) for increasing lateral compression (from left to right) from (a) top and (b) lateral views.
        Snapshots correspond to experimental surface pressure values $\pi = $1, 24, and 28 mN m$^{-1}$. These values were obtained by correlating the generalized area fraction $\zeta_{2D}$ used in simulations with that employed in surface pressure experiments.
        Color shading is used to distinguish different microgel particles. For visual purposes the box of the most compressed microgels has been replicated in the $x$ and $y$ directions parallel to the plane of the interface. Water is not shown for clarity.
        }
        \label{fig:hard_many_snaps}
\end{figure*}

\subsection{Volume fraction profiles: MD simulations}
The findings from NR experiments were further supported and validated through molecular dynamics simulations. The simulations provided additional evidence and insights into the behavior and structural changes of the microgel monolayers under compression at the interface.
Fig.~\ref{fig:hard_many_snaps} presents selected snapshots of the simulated compressed states for the hard microgels, viewed from both the top (a) and the side (b). These snapshots represent equivalent surface pressure in experiments ranging from $1$~mN~m$^{-1}$ (left) to $28$~mN~m$^{-1}$ (right), with the central snapshot corresponding to $24$~mN~m$^{-1}$.

\begin{figure}[!ht]
    \center
        \includegraphics[width=.49\textwidth]{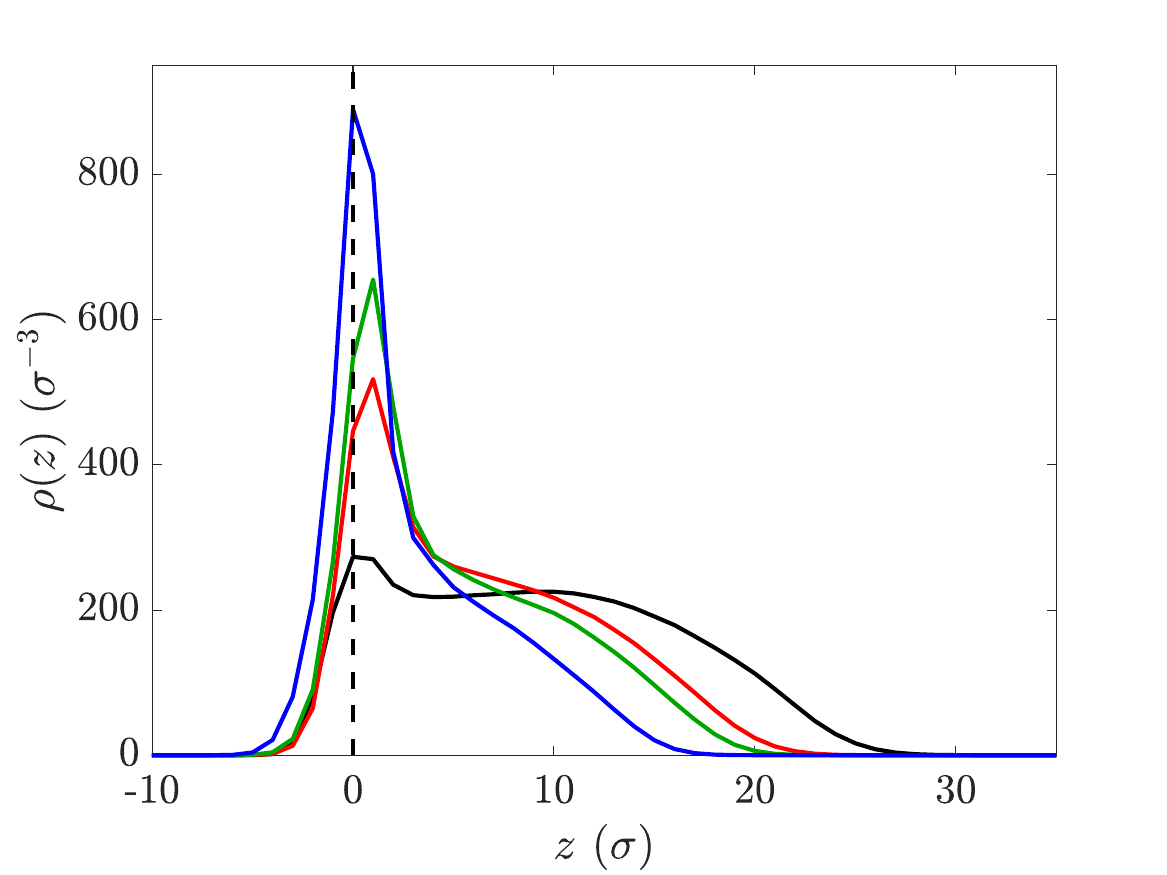}
        \caption{\footnotesize Microgel density profiles $\rho(z)$ as a function of the position along the $z$ direction computed from the simulation box containing twelve hard microgel particles at different compression, corresponding to experimental surface pressure values $\pi$= 1 (blue), 16 (green), 24 (red) and 28 (black) mNm$^{-1}$. These values were obtained by correlating the generalized area fraction $\zeta_{2D}$ used in simulations with that employed in surface pressure experiments.
        For clarity, the interface is located at z=0 and the air and water phases at $z<0$ and $z>0$ respectively.}
        \label{fig:hard_many}
\end{figure}

\begin{figure}[h!]
    \center
        \includegraphics[width=.49\textwidth]{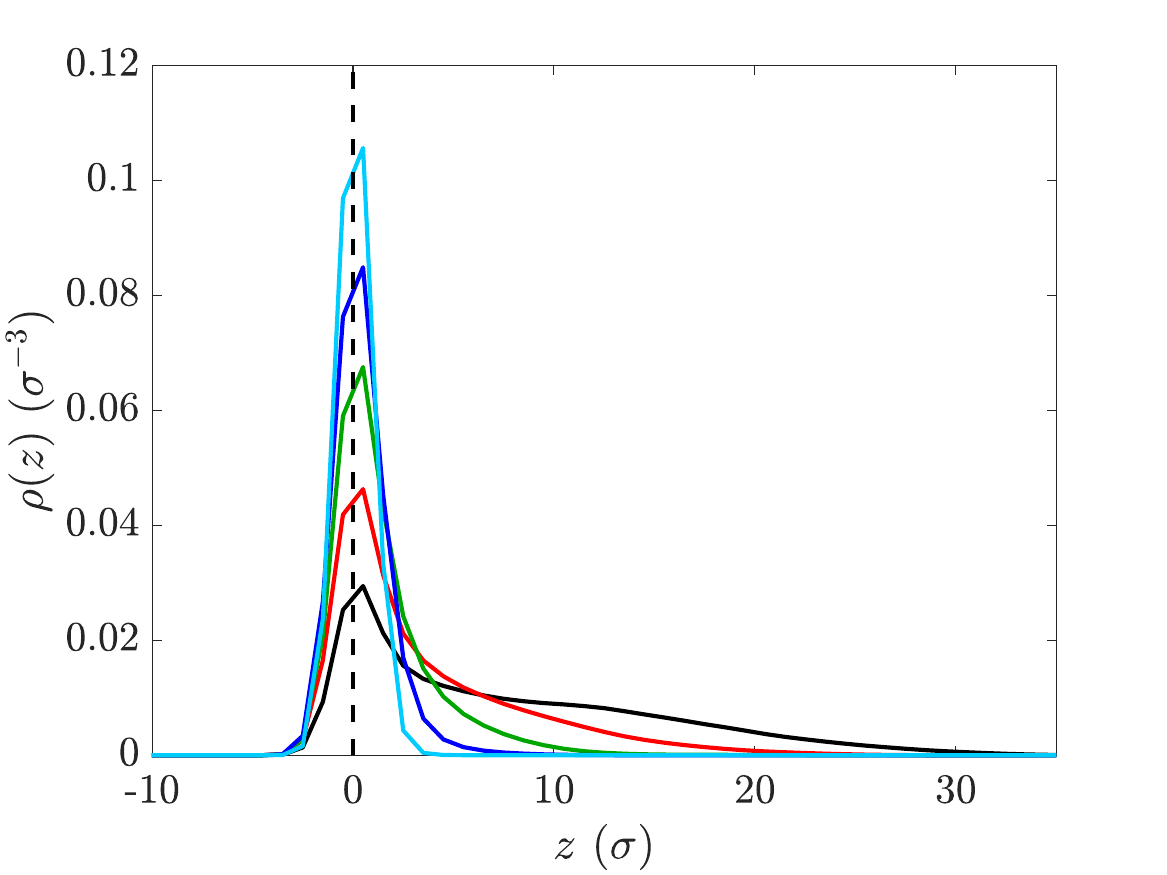}
        \caption{\footnotesize Microgel density profiles $\rho(z)$ as a function of the position along the $z$ direction computed from the simulation box containing a single soft microgel particle at different compression, corresponding to experimental surface pressure values $\pi$= 1 (cyan), 16 (blue), 25 (green), 28 (red) and 29 (black) mNm$^{-1}$. These values were obtained by correlating the generalized area fraction $\zeta_{2D}$ used in simulations with that employed in surface pressure experiments. For clarity, the interface is located at z=0 (dashed line), and the air and water phases at $z<0$ and $z>0$ respectively.}
        \label{fig:ULC_single}
\end{figure}

The density profiles depicted in Fig.~\ref{fig:hard_many} for the hard microgels exhibit a consistent overall shape reminiscent of the $\phi(z)$ profiles determined through NR experiments.
These profiles feature a denser region around the interface (at $z=0$), and protrusions extending into both the air phase ($z<0$) and the water phase ($z>0$).
As observed in the experiments, upon compression the protrusion in water gets more extended, the height of the profiles in the water phase remains almost constant while the height of the central peak, $\rho(z=0)$, and the protrusion in air, $w_r$, decrease.
The reduction of $w_r$ is also clearly visible in Fig.~\ref{fig:hard_many}.
These findings, derived from an ensemble of twelve particles, closely resemble the outcomes achieved when simulating the compression of an individual microgel particle (refer to Fig.~S5, ESI\dag).
It is worth noting that the overall shape of the profiles extracted from the analysis of the MD simulations ($\rho(z)$, Figures~\ref{fig:hard_many} and~\ref{fig:ULC_single}) and those derived from the analysis of NR data ($\phi_{\mu g}(z)$, Figures~\ref{fig:D7}b and~\ref{fig:D3}b) cannot be precisely scaled one on top of the other due to intrinsic differences between the two techniques. Briefly, the broader interfacial peak observed in MD simulations (located at $z=0$) is attributed to the coarse-grained nature of MD simulations, where the captured detail in the microgel structure is inherently larger than the molecular scale probed by NR. Similarly, the two techniques exhibit different sensitivities towards characterizing diffuse layers, i.e., interfacial films characterized by gradual and smooth changes, with MD simulations offering greater precision in determining the position in space of all utilized beads. In this context, including a gradient in the water protrusion region in the $\phi_{\mu g}(z)$ did not improve the agreement between the model reflectivity and the experimental data.

In the case of soft microgel particles, compression simulations were performed only on an individual particle because of the large system size and the corresponding high demand of computational resources (see Methods).
As shown in the snapshots (Fig.~S7, ESI\dag) and in the microgel density profiles (Fig.~\ref{fig:ULC_single}), the soft particles remained mostly localised at the interface for a broad range of compression.
The protrusion in water starts to emerge at a compression equivalent to  $\pi = 25$ mN m$^{-1}$ (green line in Fig.~\ref{fig:ULC_single}) to become more evident at larger compressions. Soft microgel particles featuring a flat oblate shape were very recently found at the interface in dodecane-water emulsions \cite{Rey2023}. This observation is fully compatible with the direct measurement of the ULC microgel shape along the vertical direction provided in the current work.
The absence of cross-linker is also responsible of the increased flattening of the particles at the interface and is consistent with previous observations \cite{Sch21, Sco19, Boc22}.

\section{Discussion}
The precise characterisation of the vertical structure in microgel monolayers is of vital importance for understanding the behaviour of stimuli-responsive microgels \cite{Rey2023}.
Such information is not directly accessed by conventional interfacial technique, while it is the main result of a specular neutron reflectometry experiment.
Indeed, the analysis of NR data allowed us to determine $\phi_{\mu g}(z)$ profiles of the volume fraction of the microgels composing the monolayer along the direction $z$ perpendicular to the interface.
The model we employed successfully captured the complex behavior of the microgels, revealing a denser region at the interface surrounded by distinct protrusions in the air and water phases.
Notably, we observed that the maximum volume fraction occupied by the microgel particles was consistently found at the air-water interface for all investigated samples.
The shape of the volume fraction profiles indicates that both soft and hard microgel particles undergo significant deformation, with many polymer chains spreading at the air-water interface due to NIPAM surface activity.
The deformation is particularly pronounced for the soft particles as they are characterised by a lower polymer density and by a larger hydration \cite{Sco19} than their hard counterpart.
As expected, $\phi_{\mu g}(0)$ initially increases upon compression for both samples.

Consistent with surface pressure measurements performed on similar systems by other groups \cite{Pin14,Pic17}, the microgel particles exhibit a transition from a gas phase to a denser phase (indicated as a liquid expanded-like state \cite{Pin14,Pic17}) as their coronas start to interact. This transition is accompanied by the emergence of a hexagonal order, as demonstrated by AFM measurements \cite{Pic17, Sco22Rev}.
Together with the lateral compression of microgels, this transition causes the observed increase in surface pressure (see Fig.~S1, ESI\dag) and in surface elasticity (Fig.~\ref{fig:Cs}).

Both NR and MD simulations indicate that, in addition to the polymer at the interface, a more hydrated polymer region is formed in the water phase. However, despite the larger hydration, the total amount of polymer in this region is 6-10 times larger than that at the interface due to its significant extension. As both hard and soft microgels are compressed in the range $1 < \pi < 13$~mN~m$^{-1}$, the overall microgel VFP increases, reflecting the greater amount of material in the probed surface area, without undergoing noticeable changes in shape as determined by the comparison of the $\phi_{\mu g}(z)$ profiles. This is in agreement with the observation that microgels, independently of their cross-linker content, interact at these compression levels through their polymer chains flattened at the interface. This results in particle coronas that can interpenetrate without inducing significant structural and morphological changes in the microgel particles.

Upon further compression ($\pi > 13$~mN~m$^{-1}$), both monolayers exhibit a peak in their surface elasticity, $\varepsilon$, which then starts to decrease while the surface pressure continues to increase, suggesting the formation of more rigid and compact film.
In the literature, this pressure range has been interpreted as the continuation of an ordered liquid phase, where the packing of microgel particles remains hexagonal but is characterized by a decreasing lattice distance \cite{Rey16, Pic17, Sco22Rev}.
Despite the apparent lack of changes in the monolayer structure in this interval of compressions, NR data indicates that the pressure at which the maximum elasticity is displayed represents a turning point for the structure of the hard microgel monolayer.
Indeed, at this compression threshold, $\phi_{\mu g}(z)$ changes significantly.
The amount of polymer at the interface levels off to a value corresponding to approximately 40\% surface coverage, representing probably the saturation limits at which polymer chains belonging to the particle corona can interpenetrate.
Simultaneously, the protrusion of polymer chains in the air drastically reduces, while it substantially increases (by 75\%) in the water phase, as also confirmed by MD simulations.
These changes are possible if the polymer chains in the particle corona can desorb from the interface to move deeper into the water phase.
As the portion of the microgel protruding in the air should not be affected by interpenetration at these compression levels, the reduction of the corresponding region in the VFP is very likely due to the vertical displacement of the entire microgel particle 4 nm deeper into the water phase.

This tendency of the hard microgels to move away from the interface can be related to their behavior once the maximum in surface compressibility $\varepsilon$ is reached.
In this condition, microgel particles are prone either to desorb into the aqueous sub-phase, forming multi-layers, or promoting the buckling of the surface \cite{Pin14}.
At the highest surface pressure utilized in the NR experiments, which is very close to the collapse of the monolayer, clear fringes appeared in the low-$Q$ portion of the data.
These are commonly associated to the beginning of the formation of a compact and thick layer, presumably indicating the onset of the transition between the liquid expanded and solid-like phases.
This behavior is driven by the interaction of the microgel particles \emph{via} their portion flattened at the interface, as indicated by the pronounced asymmetry of the volume fraction profiles and by the constant height ($A_r$) of the protrusion in the water.

The ultra-soft microgels exhibit a behavior similar to that of the hard particles in the gas phase and at the onset of the transition, particularly at surface pressures lower than the one corresponding to the maximum in surface elasticity, $\pi \approx 14$ mN~m$^{-1}$.
Here, the raise in surface pressure is caused by the interaction and interpenetration of the polymer chains at the interface.
For further compression levels, elasticity decreases, as previously observed for the hard particle monolayer.
However, the molecular reorganization responsible for this is different; for monolayers of soft microgels, a broadening of the interfacial region identified by the Gaussian function is observed, indicating that the polymer chains at the interface deform upon compression.
Despite Fig.~\ref{fig:D3}(b) shows the decrease in the maximum of $\phi_{\mu g}(z=0)$, the broadening of the interface leads to an increasing amount of polymer at the interface, as determined by \(\int_{-3w_g}^{+3w_g} \phi_{\mu g}(z) \,dz\).
Unlike the hard microgels, polymer chains located in water, just below the interface, can also interpenetrate without causing migration of some of them deeper into the water, as indicated by the constant extension of the protrusion in water found for the soft monolayer.
This enhanced interpenetration in water is consistent with the fact that the sparsely crosslinked network of swollen ultra-soft microgels present a larger mesh-size compared to crosslinked hard microgels \cite{Sco19}.

The broadening of the peak at the interface, determined experimentally to be less than 1~nm, is not visible in the density profiles obtained from the computer simulations. This is because the coarse-grained simulations were performed using particles with a size larger than the measured broadening.
However, they are able to capture the increase in the water protrusion, confirming to a good extent the validity of the experimental results.

\section{Conclusions}
\begin{figure}[!ht]
    \center
        \includegraphics[width=.49\textwidth]{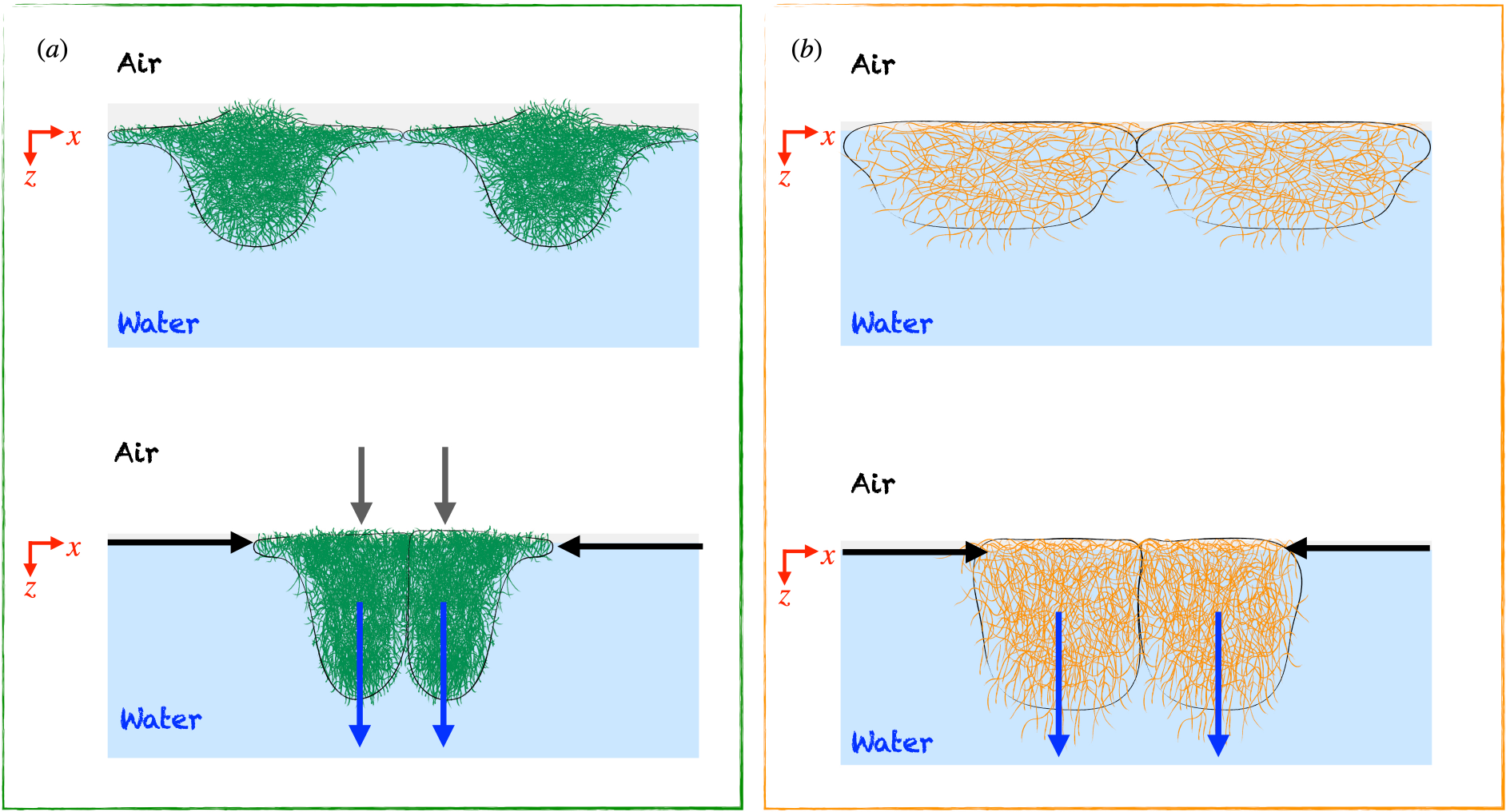}
        \caption{\footnotesize Sketch of hard (a) and soft (b) microgels pushed together at the interface.
        Black arrows represent the compression of the monolayer.
        Blue arrows represent the direction of the displacement of polymer in the aqueous sub-phase.
        The grey shaded areas represent the extension of microgels in air.
        The grey arrows represent the direction of the displacement of polymer in the air.
        }
        \label{fig:sketch}
\end{figure}

The different behavior of soft and hard microgels under compression can be attributed to their different architectures.
Cross-linked microgels are characterized by a dense core surrounded by a fuzzy corona \cite{Sti04FF}, while soft microgels synthesized without the addition of any cross-linker have a more homogeneous polymer distribution within their volume \cite{Sco19}.
These differences are preserved at the interface, where hard microgels resemble fried eggs with the hard core in their center, while soft microgels assume a pancake-like structure  \cite{Sco19, Sch21}, as depicted in the top of the sketches in Figs.~\ref{fig:sketch}(a) and~\ref{fig:sketch}(b).
The grey areas represent the protrusion of microgels in the air, as can be observed the hard microgels protrude more into air as compared to the soft ones (the proportion of the extension in air and water in the sketches are altered for clarity).

The lateral compression of hard microgels, horizontal black arrows in the bottom of panel~\ref{fig:sketch}(a), initially leads to the interaction of coronas of neighboring particles and then to the interpenetration of these regions with limited interaction of their cores. As a result, polymer chains that belong to the corona regions can rearrange, forming loops in the water, and the entire microgel particles can slip on each other, moving away from the interface.
This scenario is illustrated in the bottom of Fig.~\ref{fig:sketch}(a) where the vertical arrows show that the hard microgels are pushed more in the water (blue arrows) and away from the interface (grey arrows).

On the other hand, soft microgels do not have a significant gradient of compressibility within their volume \cite{Sch21}.
When pushed together, they can deform easily and reach a uniform coverage of the interface, as also observed in their dry state \cite{Sco19}. Yet, at one point, further compression becomes harder, and the microgels expand orthogonal to the interface, leading to an increased protrusion in water and to the broadening of the interfacial film as depicted by the vertical blue arrows in in Fig.~\ref{fig:sketch}(b).
However, the protrusion of ultra-soft microgels in the air remains almost constant as shown by the shaded gray ares in Fig.~\ref{fig:sketch}(b).

It is also noteworthy that the same level of protrusion as observed for the ULC microgels in air ($\simeq 5$~nm, Fig.~\ref{fig:D3}(b)) is only reached by the hard microgels at a higher surface pressure, specifically $\pi\gtrsim13$~mN$\cdot$m$^{-1}$ (Fig.~\ref{fig:D7}(b)).

These findings show that there is a maximum polymer density that can be allocated at the interface, independently on both the compression of the monolayer and the particle softness.
At the same time, most of the polymer is protruding into the water subphase.
As we reported in a previous study, the polymer in the water collapses onto the interface when temperature is increased above the pNIPAM volume phase transition temperature \cite{Boc22}.
However, since we show here that at different surface pressure there is a maximum occupancy of the interface, the polymer can be no longer distributed at the interface.
The collapse of polymer from the water phase onto the interface will then lead to mechanical stress in the monolayer.
In relation to emulsions, these stresses, associated with the curvature of the droplet surface, will destabilise and break the emulsion.

Thanks to our observations, we can rule out the mechanism that was originally believed being at the cause of the destabilisation: the detachment of the microgels from the interface because they become more hydrophobic above their VPTT. Our work provides experimental evidence to fully understand the stability of the microgel stabilised Pickering-like emulsions and will guide researchers from different fields to harness the power of microgels to realize sustainable smart emulsions. These materials will be pivotal in the general efforts to achieve a greener economy since smart Pickering emulsion formulations of cosmetics, detergents and shampoos can drastically reduce energy costs removing the high temperature requirements of surfactant-based formulations (cold processing)\cite{Gui21}.


Furthermore, the combination of simulations and NR experiments presented here can be of importance also to advance the rational design and consequently the use of soft colloids for biomedical applications.
For instance, recently ``synthetic antibodies'' have been prepared for a broad range of target molecules combining NIPAM and a limited set of charged and hydrophobic monomers \cite{Hau20}.
These abiotic receptors are currently being assessed as robust substitutes for antibodies in diagnostic and therapeutic applications.
The ability of these pNIPAM-based nanogels to bind multimeric receptors, along with their demonstrated equilibrium kinetics that depend on the swelling state of the microgels (i.e., its softness), have shown the potential of these materials in various application.

\section*{Author Contributions}
Y.G. Formal Analysis, Methodology, Software, Data Curation, Writing – Original Draft, Writing – Review \& Editing; F.C. Formal Analysis, Investigation, Writing – Original Draft, Writing – Review \& Editing; S.B. Formal Analysis, Investigation, Writing – Original Draft; A.M. Investigation, Writing – Review \& Editing; M.M.S. Investigation, Writing – Review \& Editing; W.R. Funding Acquisition, Writing – Review \& Editing; E.Z. Funding Acquisition, Methodology, Supervision, Writing – Original Draft, Writing – Review \& Editing; A.S. Conceptualization, Funding Acquisition, Formal Analysis, Resources, Investigation, Methodology, Supervision, Project Administration, Writing – Original Draft, Writing – Review \& Editing.

\section*{Conflicts of interest}
There are no conflicts to declare.

\section*{Acknowledgements}

\noindent AS, WR, SB and MMS thank the Deutsche Forschungsgemeinschaft within projects A3 and B8 of the SFB 985 - Functional Microgels and Microgel Systems.
EZ acknowledges support from EU MSCA Doctoral Network QLUSTER, Grant Agreement 101072964 and CINECA-ISCRA for HPC resources. AM acknowledges the financial support from MICINN under grant PID2021-129054NA-I00 and the IKUR Strategy of the Basque Government. This work was conducted within the research network Biobarriers - Health, Disorders and Healing funded by the Knowledge Foundation, Sweden (grant number 20190010). The authors thank M. Brugnoni for the microgel synthesis.

\noindent\textbf{Data availability}

\noindent The authors acknowledge the Institut Laue – Langevin for the awarded beamtime (DOIs: 10.5291/ILL-DATA.EASY-462 and 10.5291/ILL-DATA.9-11-2099) on the neutron reflectometer FIGARO.
The data generated and analysed during this study are openly available in RADAR4Chem at https://doi.org/10.22000/1747.
The software utilised for the analysis of neutron reflectometry data (Aurore - VFP version) is available at \url{https://sourceforge.net/projects/aurorenr/}.

\includepdf[pages=-]{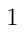}

\balance

\bibliography{refs}

\providecommand{\latin}[1]{#1}
\makeatletter
\providecommand{\doi}
  {\begingroup\let\do\@makeother\dospecials
  \catcode`\{=1 \catcode`\}=2 \doi@aux}
\providecommand{\doi@aux}[1]{\endgroup\texttt{#1}}
\makeatother
\providecommand*\mcitethebibliography{\thebibliography}
\csname @ifundefined\endcsname{endmcitethebibliography}
  {\let\endmcitethebibliography\endthebibliography}{}
\begin{mcitethebibliography}{78}
\providecommand*\natexlab[1]{#1}
\providecommand*\mciteSetBstSublistMode[1]{}
\providecommand*\mciteSetBstMaxWidthForm[2]{}
\providecommand*\mciteBstWouldAddEndPuncttrue
  {\def\EndOfBibitem{\unskip.}}
\providecommand*\mciteBstWouldAddEndPunctfalse
  {\let\EndOfBibitem\relax}
\providecommand*\mciteSetBstMidEndSepPunct[3]{}
\providecommand*\mciteSetBstSublistLabelBeginEnd[3]{}
\providecommand*\EndOfBibitem{}
\mciteSetBstSublistMode{f}
\mciteSetBstMaxWidthForm{subitem}{(\alph{mcitesubitemcount})}
\mciteSetBstSublistLabelBeginEnd
  {\mcitemaxwidthsubitemform\space}
  {\relax}
  {\relax}

\bibitem[Binks(2002)]{Bin02}
Binks,~B.~P. Particles as surfactants—similarities and differences.
  \emph{Current opinion in colloid \& interface science} \textbf{2002},
  \emph{7}, 21--41\relax
\mciteBstWouldAddEndPuncttrue
\mciteSetBstMidEndSepPunct{\mcitedefaultmidpunct}
{\mcitedefaultendpunct}{\mcitedefaultseppunct}\relax
\EndOfBibitem
\bibitem[Dekker \latin{et~al.}(2023)Dekker, Velandia, Kibbelaar, Morcy,
  Sadtler, Roques-Carmes, Groenewold, Kegel, Velikov, and Bonn]{Dek23}
Dekker,~R.~I.; Velandia,~S.~F.; Kibbelaar,~H.~V.; Morcy,~A.; Sadtler,~V.;
  Roques-Carmes,~T.; Groenewold,~J.; Kegel,~W.~K.; Velikov,~K.~P.; Bonn,~D. Is
  there a difference between surfactant-stabilised and Pickering emulsions?
  \emph{Soft Matter} \textbf{2023}, \emph{19}, 1941--1951\relax
\mciteBstWouldAddEndPuncttrue
\mciteSetBstMidEndSepPunct{\mcitedefaultmidpunct}
{\mcitedefaultendpunct}{\mcitedefaultseppunct}\relax
\EndOfBibitem
\bibitem[Singh and Sarkar(2011)Singh, and Sarkar]{Sin11}
Singh,~H.; Sarkar,~A. Behaviour of protein-stabilised emulsions under various
  physiological conditions. \emph{Advances in colloid and interface science}
  \textbf{2011}, \emph{165}, 47--57\relax
\mciteBstWouldAddEndPuncttrue
\mciteSetBstMidEndSepPunct{\mcitedefaultmidpunct}
{\mcitedefaultendpunct}{\mcitedefaultseppunct}\relax
\EndOfBibitem
\bibitem[Hobson \latin{et~al.}(2018)Hobson, Edwards, Slater, Martin, Owen, and
  Rannard]{Hob18}
Hobson,~J.~J.; Edwards,~S.; Slater,~R.~A.; Martin,~P.; Owen,~A.; Rannard,~S.~P.
  Branched copolymer-stabilised nanoemulsions as new candidate oral drug
  delivery systems. \emph{RSC advances} \textbf{2018}, \emph{8},
  12984--12991\relax
\mciteBstWouldAddEndPuncttrue
\mciteSetBstMidEndSepPunct{\mcitedefaultmidpunct}
{\mcitedefaultendpunct}{\mcitedefaultseppunct}\relax
\EndOfBibitem
\bibitem[Ramsden(1904)]{Ram04}
Ramsden,~W. Separation of solids in the surface-layers of solutions and
  ‘suspensions’(observations on surface-membranes, bubbles, emulsions, and
  mechanical coagulation).—Preliminary account. \emph{Proceedings of the
  royal Society of London} \textbf{1904}, \emph{72}, 156--164\relax
\mciteBstWouldAddEndPuncttrue
\mciteSetBstMidEndSepPunct{\mcitedefaultmidpunct}
{\mcitedefaultendpunct}{\mcitedefaultseppunct}\relax
\EndOfBibitem
\bibitem[Pickering(1907)]{Pic07}
Pickering,~S.~U. Emulsions. \emph{Journal of the Chemical Society,
  Transactions} \textbf{1907}, \emph{91}, 2001--2021\relax
\mciteBstWouldAddEndPuncttrue
\mciteSetBstMidEndSepPunct{\mcitedefaultmidpunct}
{\mcitedefaultendpunct}{\mcitedefaultseppunct}\relax
\EndOfBibitem
\bibitem[Hossain \latin{et~al.}(2021)Hossain, Deeming, and Edler]{Hos21}
Hossain,~K. M.~Z.; Deeming,~L.; Edler,~K.~J. Recent progress in Pickering
  emulsions stabilised by bioderived particles. \emph{RSC advances}
  \textbf{2021}, \emph{11}, 39027--39044\relax
\mciteBstWouldAddEndPuncttrue
\mciteSetBstMidEndSepPunct{\mcitedefaultmidpunct}
{\mcitedefaultendpunct}{\mcitedefaultseppunct}\relax
\EndOfBibitem
\bibitem[Richtering(2012)]{Ric12}
Richtering,~W. Responsive emulsions stabilized by stimuli-sensitive microgels:
  emulsions with special non-Pickering properties. \emph{Langmuir}
  \textbf{2012}, \emph{28}, 17218--17229\relax
\mciteBstWouldAddEndPuncttrue
\mciteSetBstMidEndSepPunct{\mcitedefaultmidpunct}
{\mcitedefaultendpunct}{\mcitedefaultseppunct}\relax
\EndOfBibitem
\bibitem[Destribats \latin{et~al.}(2011)Destribats, Lapeyre, Wolfs, Sellier,
  Leal-Calderon, Ravaine, and Schmitt]{Des11}
Destribats,~M.; Lapeyre,~V.; Wolfs,~M.; Sellier,~E.; Leal-Calderon,~F.;
  Ravaine,~V.; Schmitt,~V. Soft microgels as Pickering emulsion stabilisers:
  role of particle deformability. \emph{Soft Matter} \textbf{2011}, \emph{7},
  7689--7698\relax
\mciteBstWouldAddEndPuncttrue
\mciteSetBstMidEndSepPunct{\mcitedefaultmidpunct}
{\mcitedefaultendpunct}{\mcitedefaultseppunct}\relax
\EndOfBibitem
\bibitem[Kwok \latin{et~al.}(2019)Kwok, Sun, and Ngai]{Kwo19}
Kwok,~M.-h.; Sun,~G.; Ngai,~T. Microgel particles at interfaces: phenomena,
  principles, and opportunities in food sciences. \emph{Langmuir}
  \textbf{2019}, \emph{35}, 4205--4217\relax
\mciteBstWouldAddEndPuncttrue
\mciteSetBstMidEndSepPunct{\mcitedefaultmidpunct}
{\mcitedefaultendpunct}{\mcitedefaultseppunct}\relax
\EndOfBibitem
\bibitem[Pinaud \latin{et~al.}(2014)Pinaud, Geisel, Mass{\'{e}}, Catargi, Isa,
  Richtering, Ravaine, and Schmitt]{Pin14}
Pinaud,~F.; Geisel,~K.; Mass{\'{e}},~P.; Catargi,~B.; Isa,~L.; Richtering,~W.;
  Ravaine,~V.; Schmitt,~V. Adsorption of microgels at an oil{\textendash}water
  interface: correlation between packing and 2D elasticity. \emph{Soft Matter}
  \textbf{2014}, \emph{10}, 6963--6974\relax
\mciteBstWouldAddEndPuncttrue
\mciteSetBstMidEndSepPunct{\mcitedefaultmidpunct}
{\mcitedefaultendpunct}{\mcitedefaultseppunct}\relax
\EndOfBibitem
\bibitem[Scotti \latin{et~al.}(2022)Scotti, Schulte, Lopez, Crassous, Bochenek,
  and Richtering]{Sco22Rev}
Scotti,~A.; Schulte,~M.~F.; Lopez,~C.~G.; Crassous,~J.~J.; Bochenek,~S.;
  Richtering,~W. How softness matters in soft nanogels and nanogel assemblies.
  \emph{Chemical Reviews} \textbf{2022}, \emph{122}, 11675--11700\relax
\mciteBstWouldAddEndPuncttrue
\mciteSetBstMidEndSepPunct{\mcitedefaultmidpunct}
{\mcitedefaultendpunct}{\mcitedefaultseppunct}\relax
\EndOfBibitem
\bibitem[Bochenek \latin{et~al.}(2022)Bochenek, Camerin, Zaccarelli, Maestro,
  Schmidt, Richtering, and Scotti]{Boc22}
Bochenek,~S.; Camerin,~F.; Zaccarelli,~E.; Maestro,~A.; Schmidt,~M.~M.;
  Richtering,~W.; Scotti,~A. In-situ study of the impact of temperature and
  architecture on the interfacial structure of microgels. \emph{Nature
  Communications} \textbf{2022}, \emph{13}, 3744\relax
\mciteBstWouldAddEndPuncttrue
\mciteSetBstMidEndSepPunct{\mcitedefaultmidpunct}
{\mcitedefaultendpunct}{\mcitedefaultseppunct}\relax
\EndOfBibitem
\bibitem[Schulte \latin{et~al.}(2021)Schulte, Bochenek, Brugnoni, Scotti,
  Mourran, and Richtering]{Sch21}
Schulte,~M.~F.; Bochenek,~S.; Brugnoni,~M.; Scotti,~A.; Mourran,~A.;
  Richtering,~W. Stiffness tomography of ultra-soft nanogels by atomic force
  microscopy. \emph{Angewandte Chemie International Edition} \textbf{2021},
  \emph{60}, 2280--2287\relax
\mciteBstWouldAddEndPuncttrue
\mciteSetBstMidEndSepPunct{\mcitedefaultmidpunct}
{\mcitedefaultendpunct}{\mcitedefaultseppunct}\relax
\EndOfBibitem
\bibitem[Houston \latin{et~al.}(2022)Houston, Fruhner, de~la Cotte,
  Rojo~Gonz{\'a}lez, Petrunin, Gasser, Schweins, Allgaier, Richtering,
  Fernandez-Nieves, \latin{et~al.} others]{Hou22}
Houston,~J.~E.; Fruhner,~L.; de~la Cotte,~A.; Rojo~Gonz{\'a}lez,~J.;
  Petrunin,~A.~V.; Gasser,~U.; Schweins,~R.; Allgaier,~J.; Richtering,~W.;
  Fernandez-Nieves,~A.; others Resolving the different bulk moduli within
  individual soft nanogels using small-angle neutron scattering. \emph{Science
  Advances} \textbf{2022}, \emph{8}, eabn6129\relax
\mciteBstWouldAddEndPuncttrue
\mciteSetBstMidEndSepPunct{\mcitedefaultmidpunct}
{\mcitedefaultendpunct}{\mcitedefaultseppunct}\relax
\EndOfBibitem
\bibitem[Rey \latin{et~al.}(2016)Rey, Fern{\'a}ndez-Rodr{\'\i}guez, Steinacher,
  Scheidegger, Geisel, Richtering, Squires, and Isa]{Rey16}
Rey,~M.; Fern{\'a}ndez-Rodr{\'\i}guez,~M.~{\'A}.; Steinacher,~M.;
  Scheidegger,~L.; Geisel,~K.; Richtering,~W.; Squires,~T.~M.; Isa,~L.
  Isostructural solid--solid phase transition in monolayers of soft core--shell
  particles at fluid interfaces: structure and mechanics. \emph{Soft Matter}
  \textbf{2016}, \emph{12}, 3545--3557\relax
\mciteBstWouldAddEndPuncttrue
\mciteSetBstMidEndSepPunct{\mcitedefaultmidpunct}
{\mcitedefaultendpunct}{\mcitedefaultseppunct}\relax
\EndOfBibitem
\bibitem[Brugger \latin{et~al.}(2010)Brugger, Vermant, and Richtering]{Bru10}
Brugger,~B.; Vermant,~J.; Richtering,~W. Interfacial layers of
  stimuli-responsive poly-(N-isopropylacrylamide-co-methacrylicacid)
  ({PNIPAM}-co-{MAA}) microgels characterized by interfacial rheology and
  compression isotherms. \emph{Physical Chemistry Chemical Physics}
  \textbf{2010}, \emph{12}, 14573--14578\relax
\mciteBstWouldAddEndPuncttrue
\mciteSetBstMidEndSepPunct{\mcitedefaultmidpunct}
{\mcitedefaultendpunct}{\mcitedefaultseppunct}\relax
\EndOfBibitem
\bibitem[Akentiev \latin{et~al.}(2017)Akentiev, Rybnikova, Novikova, Timoshen,
  Zorin, and Noskov]{Ake17}
Akentiev,~A.~V.; Rybnikova,~G.~S.; Novikova,~A.~A.; Timoshen,~K.~A.;
  Zorin,~I.~M.; Noskov,~B.~A. Dynamic elasticity of films formed by
  poly(N-isopropylacrylamide) microparticles on a water surface. \emph{Colloid
  Journal} \textbf{2017}, \emph{79}, 571--576\relax
\mciteBstWouldAddEndPuncttrue
\mciteSetBstMidEndSepPunct{\mcitedefaultmidpunct}
{\mcitedefaultendpunct}{\mcitedefaultseppunct}\relax
\EndOfBibitem
\bibitem[Tatry \latin{et~al.}(2023)Tatry, Laurichesse, Vermant, Ravaine, and
  Schmitt]{Tat23}
Tatry,~M.-C.; Laurichesse,~E.; Vermant,~J.; Ravaine,~V.; Schmitt,~V.
  Interfacial rheology of model water–air microgels laden interfaces: Effect
  of cross-linking. \emph{Journal of Colloid and Interface Science}
  \textbf{2023}, \emph{629}, 288--299\relax
\mciteBstWouldAddEndPuncttrue
\mciteSetBstMidEndSepPunct{\mcitedefaultmidpunct}
{\mcitedefaultendpunct}{\mcitedefaultseppunct}\relax
\EndOfBibitem
\bibitem[Schmidt \latin{et~al.}(2023)Schmidt, Ruiz-Franco, Bochenek, Camerin,
  Zaccarelli, and Scotti]{Sch23}
Schmidt,~M.~M.; Ruiz-Franco,~J.; Bochenek,~S.; Camerin,~F.; Zaccarelli,~E.;
  Scotti,~A. Interfacial fluid rheology of soft particles. \emph{Physical
  Review Letters} \textbf{2023}, \emph{131}, 258202\relax
\mciteBstWouldAddEndPuncttrue
\mciteSetBstMidEndSepPunct{\mcitedefaultmidpunct}
{\mcitedefaultendpunct}{\mcitedefaultseppunct}\relax
\EndOfBibitem
\bibitem[Kawamoto \latin{et~al.}(2023)Kawamoto, Yanagi, Nishizawa, Minato, and
  Suzuki]{Kaw23}
Kawamoto,~T.; Yanagi,~K.; Nishizawa,~Y.; Minato,~H.; Suzuki,~D. The compression
  of deformed microgels at an air/water interface. \emph{Chemical
  Communications} \textbf{2023}, \emph{59}, 13289–13292\relax
\mciteBstWouldAddEndPuncttrue
\mciteSetBstMidEndSepPunct{\mcitedefaultmidpunct}
{\mcitedefaultendpunct}{\mcitedefaultseppunct}\relax
\EndOfBibitem
\bibitem[Rey \latin{et~al.}(2023)Rey, Kolker, Richards, Malhotra, Glen, Li,
  Laidlaw, Renggli, Vermant, Schofield, Fujii, L\"{o}wen, and Clegg]{Rey2023}
Rey,~M.; Kolker,~J.; Richards,~J.~A.; Malhotra,~I.; Glen,~T.~S.; Li,~N. Y.~D.;
  Laidlaw,~F. H.~J.; Renggli,~D.; Vermant,~J.; Schofield,~A.~B.; Fujii,~S.;
  L\"{o}wen,~H.; Clegg,~P.~S. Interactions between interfaces dictate
  stimuli-responsive emulsion behaviour. \emph{Nature Communications}
  \textbf{2023}, \emph{14}, 6723\relax
\mciteBstWouldAddEndPuncttrue
\mciteSetBstMidEndSepPunct{\mcitedefaultmidpunct}
{\mcitedefaultendpunct}{\mcitedefaultseppunct}\relax
\EndOfBibitem
\bibitem[Zhang and Pelton(1999)Zhang, and Pelton]{Zhang1999}
Zhang,~J.; Pelton,~R. Poly(N-isopropylacrylamide) Microgels at the Air-Water
  Interface. \emph{Langmuir} \textbf{1999}, \emph{15}, 8032–8036\relax
\mciteBstWouldAddEndPuncttrue
\mciteSetBstMidEndSepPunct{\mcitedefaultmidpunct}
{\mcitedefaultendpunct}{\mcitedefaultseppunct}\relax
\EndOfBibitem
\bibitem[Saunders(2004)]{Saunders2004}
Saunders,~B.~R. On the Structure of Poly(N-isopropylacrylamide) Microgel
  Particles. \emph{Langmuir} \textbf{2004}, \emph{20}, 3925--3932\relax
\mciteBstWouldAddEndPuncttrue
\mciteSetBstMidEndSepPunct{\mcitedefaultmidpunct}
{\mcitedefaultendpunct}{\mcitedefaultseppunct}\relax
\EndOfBibitem
\bibitem[Nishizawa \latin{et~al.}(2019)Nishizawa, Matsui, Urayama, Kureha,
  Shibayama, Uchihashi, and Suzuki]{Nishizawa2019}
Nishizawa,~Y.; Matsui,~S.; Urayama,~K.; Kureha,~T.; Shibayama,~M.;
  Uchihashi,~T.; Suzuki,~D. Non‐Thermoresponsive Decanano‐sized Domains in
  Thermoresponsive Hydrogel Microspheres Revealed by Temperature‐Controlled
  High‐Speed Atomic Force Microscopy. \emph{Angewandte Chemie International
  Edition} \textbf{2019}, \emph{58}, 8809–8813\relax
\mciteBstWouldAddEndPuncttrue
\mciteSetBstMidEndSepPunct{\mcitedefaultmidpunct}
{\mcitedefaultendpunct}{\mcitedefaultseppunct}\relax
\EndOfBibitem
\bibitem[Nishizawa \latin{et~al.}(2021)Nishizawa, Honda, and
  Suzuki]{Nishizawa2021}
Nishizawa,~Y.; Honda,~K.; Suzuki,~D. Recent Development in the Visualization of
  Microgels. \emph{Chemistry Letters} \textbf{2021}, \emph{50},
  1226–1235\relax
\mciteBstWouldAddEndPuncttrue
\mciteSetBstMidEndSepPunct{\mcitedefaultmidpunct}
{\mcitedefaultendpunct}{\mcitedefaultseppunct}\relax
\EndOfBibitem
\bibitem[McPhee \latin{et~al.}(1993)McPhee, Tam, and Pelton]{McP93}
McPhee,~W.; Tam,~K.~C.; Pelton,~R. Poly (N-isopropylacrylamide) latices
  prepared with sodium dodecyl sulfate. \emph{Journal of colloid and interface
  science} \textbf{1993}, \emph{156}, 24--30\relax
\mciteBstWouldAddEndPuncttrue
\mciteSetBstMidEndSepPunct{\mcitedefaultmidpunct}
{\mcitedefaultendpunct}{\mcitedefaultseppunct}\relax
\EndOfBibitem
\bibitem[Bachman \latin{et~al.}(2015)Bachman, Brown, Clarke, Dhada, Douglas,
  Hansen, Herman, Hyatt, Kodlekere, Meng, \latin{et~al.} others]{Bac15}
Bachman,~H.; Brown,~A.~C.; Clarke,~K.~C.; Dhada,~K.~S.; Douglas,~A.;
  Hansen,~C.~E.; Herman,~E.; Hyatt,~J.~S.; Kodlekere,~P.; Meng,~Z.; others
  Ultrasoft, highly deformable microgels. \emph{Soft Matter} \textbf{2015},
  \emph{11}, 2018--2028\relax
\mciteBstWouldAddEndPuncttrue
\mciteSetBstMidEndSepPunct{\mcitedefaultmidpunct}
{\mcitedefaultendpunct}{\mcitedefaultseppunct}\relax
\EndOfBibitem
\bibitem[Brugnoni \latin{et~al.}(2019)Brugnoni, Nickel, Kr{\"o}ger, Scotti,
  Pich, Leonhard, and Richtering]{Bru19}
Brugnoni,~M.; Nickel,~A.~C.; Kr{\"o}ger,~L.~C.; Scotti,~A.; Pich,~A.;
  Leonhard,~K.; Richtering,~W. Synthesis and structure of deuterated ultra-low
  cross-linked poly (N-isopropylacrylamide) microgels. \emph{Polymer Chemistry}
  \textbf{2019}, \emph{10}, 2397--2405\relax
\mciteBstWouldAddEndPuncttrue
\mciteSetBstMidEndSepPunct{\mcitedefaultmidpunct}
{\mcitedefaultendpunct}{\mcitedefaultseppunct}\relax
\EndOfBibitem
\bibitem[Scotti \latin{et~al.}(2019)Scotti, Bochenek, Brugnoni,
  Fernandez-Rodriguez, Schulte, Houston, Gelissen, Potemkin, Isa, and
  Richtering]{Sco19}
Scotti,~A.; Bochenek,~S.; Brugnoni,~M.; Fernandez-Rodriguez,~M.-A.;
  Schulte,~M.~F.; Houston,~J.; Gelissen,~A.~P.; Potemkin,~I.~I.; Isa,~L.;
  Richtering,~W. Exploring the colloid-to-polymer transition for ultra-low
  crosslinked microgels from three to two dimensions. \emph{Nature
  Communications} \textbf{2019}, \emph{10}, 1418\relax
\mciteBstWouldAddEndPuncttrue
\mciteSetBstMidEndSepPunct{\mcitedefaultmidpunct}
{\mcitedefaultendpunct}{\mcitedefaultseppunct}\relax
\EndOfBibitem
\bibitem[Petrunin \latin{et~al.}(2023)Petrunin, Bochenek, Richtering, and
  Scotti]{Pet23}
Petrunin,~A.~V.; Bochenek,~S.; Richtering,~W.; Scotti,~A. Harnessing the
  polymer-particle duality of ultra-soft nanogels to stabilise smart emulsions.
  \emph{Physical Chemistry Chemical Physics} \textbf{2023}, \emph{25},
  2810--2820\relax
\mciteBstWouldAddEndPuncttrue
\mciteSetBstMidEndSepPunct{\mcitedefaultmidpunct}
{\mcitedefaultendpunct}{\mcitedefaultseppunct}\relax
\EndOfBibitem
\bibitem[Nussbaum \latin{et~al.}(2022)Nussbaum, Bergfreund, Vialetto, Isa, and
  Fischer]{Nus22}
Nussbaum,~N.; Bergfreund,~J.; Vialetto,~J.; Isa,~L.; Fischer,~P. Microgels as
  globular protein model systems. \emph{Colloids and Surfaces B: Biointerfaces}
  \textbf{2022}, \emph{217}, 112595\relax
\mciteBstWouldAddEndPuncttrue
\mciteSetBstMidEndSepPunct{\mcitedefaultmidpunct}
{\mcitedefaultendpunct}{\mcitedefaultseppunct}\relax
\EndOfBibitem
\bibitem[Tein \latin{et~al.}(2020)Tein, Zhang, and Wagner]{Tei20}
Tein,~Y.~S.; Zhang,~Z.; Wagner,~N.~J. Competitive Surface Activity of
  Monoclonal Antibodies and Nonionic Surfactants at the Air-Water Interface
  Determined by Interfacial Rheology and Neutron Reflectometry. \emph{Langmuir}
  \textbf{2020}, \emph{36}, 7814--7823\relax
\mciteBstWouldAddEndPuncttrue
\mciteSetBstMidEndSepPunct{\mcitedefaultmidpunct}
{\mcitedefaultendpunct}{\mcitedefaultseppunct}\relax
\EndOfBibitem
\bibitem[Wood \latin{et~al.}(2023)Wood, Razinkov, Qi, Roberts, Vermant, and
  Furst]{Woo23}
Wood,~C.~V.; Razinkov,~V.~I.; Qi,~W.; Roberts,~C.~J.; Vermant,~J.; Furst,~E.~M.
  Antibodies Adsorbed to the Air-Water Interface Form Soft Glasses.
  \emph{Langmuir} \textbf{2023}, \emph{39}, 7775--7782\relax
\mciteBstWouldAddEndPuncttrue
\mciteSetBstMidEndSepPunct{\mcitedefaultmidpunct}
{\mcitedefaultendpunct}{\mcitedefaultseppunct}\relax
\EndOfBibitem
\bibitem[Zahn \latin{et~al.}(1999)Zahn, Lenke, and Maret]{Zah99}
Zahn,~K.; Lenke,~R.; Maret,~G. Two-stage melting of paramagnetic colloidal
  crystals in two dimensions. \emph{Physical review letters} \textbf{1999},
  \emph{82}, 2721\relax
\mciteBstWouldAddEndPuncttrue
\mciteSetBstMidEndSepPunct{\mcitedefaultmidpunct}
{\mcitedefaultendpunct}{\mcitedefaultseppunct}\relax
\EndOfBibitem
\bibitem[Zahn and Maret(2000)Zahn, and Maret]{Zah00}
Zahn,~K.; Maret,~G. Dynamic criteria for melting in two dimensions.
  \emph{Physical Review Letters} \textbf{2000}, \emph{85}, 3656\relax
\mciteBstWouldAddEndPuncttrue
\mciteSetBstMidEndSepPunct{\mcitedefaultmidpunct}
{\mcitedefaultendpunct}{\mcitedefaultseppunct}\relax
\EndOfBibitem
\bibitem[Kelleher \latin{et~al.}(2017)Kelleher, Guerra, Hollingsworth, and
  Chaikin]{Kel17}
Kelleher,~C.~P.; Guerra,~R.~E.; Hollingsworth,~A.~D.; Chaikin,~P.~M. Phase
  behavior of charged colloids at a fluid interface. \emph{Phys. Rev. E}
  \textbf{2017}, \emph{95}, 022602\relax
\mciteBstWouldAddEndPuncttrue
\mciteSetBstMidEndSepPunct{\mcitedefaultmidpunct}
{\mcitedefaultendpunct}{\mcitedefaultseppunct}\relax
\EndOfBibitem
\bibitem[Deshmukh \latin{et~al.}(2014)Deshmukh, Maestro, Duits, van~den Ende,
  Stuart, and Mugele]{Des14}
Deshmukh,~O.~S.; Maestro,~A.; Duits,~M.~H.; van~den Ende,~D.; Stuart,~M.~C.;
  Mugele,~F. Equation of state and adsorption dynamics of soft microgel
  particles at an air--water interface. \emph{Soft matter} \textbf{2014},
  \emph{10}, 7045--7050\relax
\mciteBstWouldAddEndPuncttrue
\mciteSetBstMidEndSepPunct{\mcitedefaultmidpunct}
{\mcitedefaultendpunct}{\mcitedefaultseppunct}\relax
\EndOfBibitem
\bibitem[Scheidegger \latin{et~al.}(2017)Scheidegger,
  Fern{\'a}ndez-Rodr{\'\i}guez, Geisel, Zanini, Elnathan, Richtering, and
  Isa]{Sch17}
Scheidegger,~L.; Fern{\'a}ndez-Rodr{\'\i}guez,~M.~{\'A}.; Geisel,~K.;
  Zanini,~M.; Elnathan,~R.; Richtering,~W.; Isa,~L. Compression and deposition
  of microgel monolayers from fluid interfaces: particle size effects on
  interface microstructure and nanolithography. \emph{Physical Chemistry
  Chemical Physics} \textbf{2017}, \emph{19}, 8671--8680\relax
\mciteBstWouldAddEndPuncttrue
\mciteSetBstMidEndSepPunct{\mcitedefaultmidpunct}
{\mcitedefaultendpunct}{\mcitedefaultseppunct}\relax
\EndOfBibitem
\bibitem[Schmidt \latin{et~al.}(2010)Schmidt, Zeiser, Hellweg, Duschl, Fery,
  and M{\"o}hwald]{Sch10}
Schmidt,~S.; Zeiser,~M.; Hellweg,~T.; Duschl,~C.; Fery,~A.; M{\"o}hwald,~H.
  Adhesion and mechanical properties of PNIPAM microgel films and their
  potential use as switchable cell culture substrates. \emph{Advanced
  Functional Materials} \textbf{2010}, \emph{20}, 3235--3243\relax
\mciteBstWouldAddEndPuncttrue
\mciteSetBstMidEndSepPunct{\mcitedefaultmidpunct}
{\mcitedefaultendpunct}{\mcitedefaultseppunct}\relax
\EndOfBibitem
\bibitem[Cors \latin{et~al.}(2017)Cors, Wrede, Genix, Anselmetti, Oberdisse,
  and Hellweg]{Cor17}
Cors,~M.; Wrede,~O.; Genix,~A.-C.; Anselmetti,~D.; Oberdisse,~J.; Hellweg,~T.
  Core--shell microgel-based surface coatings with linear thermoresponse.
  \emph{Langmuir} \textbf{2017}, \emph{33}, 6804--6811\relax
\mciteBstWouldAddEndPuncttrue
\mciteSetBstMidEndSepPunct{\mcitedefaultmidpunct}
{\mcitedefaultendpunct}{\mcitedefaultseppunct}\relax
\EndOfBibitem
\bibitem[Geisel \latin{et~al.}(2014)Geisel, Isa, and Richtering]{Gei14}
Geisel,~K.; Isa,~L.; Richtering,~W. The Compressibility of pH-Sensitive
  Microgels at the Oil--Water Interface: Higher Charge Leads to Less Repulsion.
  \emph{Angewandte Chemie} \textbf{2014}, \emph{126}, 5005--5009\relax
\mciteBstWouldAddEndPuncttrue
\mciteSetBstMidEndSepPunct{\mcitedefaultmidpunct}
{\mcitedefaultendpunct}{\mcitedefaultseppunct}\relax
\EndOfBibitem
\bibitem[Geisel \latin{et~al.}(2014)Geisel, Richtering, and Isa]{Gei14b}
Geisel,~K.; Richtering,~W.; Isa,~L. Highly ordered 2D microgel arrays:
  compression versus self-assembly. \emph{Soft Matter} \textbf{2014},
  \emph{10}, 7968--7976\relax
\mciteBstWouldAddEndPuncttrue
\mciteSetBstMidEndSepPunct{\mcitedefaultmidpunct}
{\mcitedefaultendpunct}{\mcitedefaultseppunct}\relax
\EndOfBibitem
\bibitem[Kuk \latin{et~al.}(2023)Kuk, Abgarjan, Gregel, Zhou, Fadanelli,
  Buttinoni, and Karg]{Kuk23}
Kuk,~K.; Abgarjan,~V.; Gregel,~L.; Zhou,~Y.; Fadanelli,~V.~C.; Buttinoni,~I.;
  Karg,~M. Compression of colloidal monolayers at liquid interfaces: in situ
  vs. ex situ investigation. \emph{Soft Matter} \textbf{2023}, \emph{19},
  175--188\relax
\mciteBstWouldAddEndPuncttrue
\mciteSetBstMidEndSepPunct{\mcitedefaultmidpunct}
{\mcitedefaultendpunct}{\mcitedefaultseppunct}\relax
\EndOfBibitem
\bibitem[Zieli{\'n}ska \latin{et~al.}(2016)Zieli{\'n}ska, Sun, Campbell,
  Zarbakhsh, and Resmini]{Zie16}
Zieli{\'n}ska,~K.; Sun,~H.; Campbell,~R.~A.; Zarbakhsh,~A.; Resmini,~M. Smart
  nanogels at the air/water interface: structural studies by neutron
  reflectivity. \emph{Nanoscale} \textbf{2016}, \emph{8}, 4951--4960\relax
\mciteBstWouldAddEndPuncttrue
\mciteSetBstMidEndSepPunct{\mcitedefaultmidpunct}
{\mcitedefaultendpunct}{\mcitedefaultseppunct}\relax
\EndOfBibitem
\bibitem[Rey \latin{et~al.}(2020)Rey, Fernandez-Rodriguez, Karg, Isa, and
  Vogel]{Rey2020}
Rey,~M.; Fernandez-Rodriguez,~M.~A.; Karg,~M.; Isa,~L.; Vogel,~N.
  Poly-N-isopropylacrylamide Nanogels and Microgels at Fluid Interfaces.
  \emph{Accounts of Chemical Research} \textbf{2020}, \emph{53},
  414–424\relax
\mciteBstWouldAddEndPuncttrue
\mciteSetBstMidEndSepPunct{\mcitedefaultmidpunct}
{\mcitedefaultendpunct}{\mcitedefaultseppunct}\relax
\EndOfBibitem
\bibitem[Bochenek \latin{et~al.}(2021)Bochenek, Scotti, and Richtering]{Boc21}
Bochenek,~S.; Scotti,~A.; Richtering,~W. Temperature-sensitive soft microgels
  at interfaces: air--water versus oil--water. \emph{Soft Matter}
  \textbf{2021}, \emph{17}, 976--988\relax
\mciteBstWouldAddEndPuncttrue
\mciteSetBstMidEndSepPunct{\mcitedefaultmidpunct}
{\mcitedefaultendpunct}{\mcitedefaultseppunct}\relax
\EndOfBibitem
\bibitem[Stieger \latin{et~al.}(2004)Stieger, Richtering, Pedersen, and
  Lindner]{Sti04FF}
Stieger,~M.; Richtering,~W.; Pedersen,~J.~S.; Lindner,~P. Small-angle neutron
  scattering study of structural changes in temperature sensitive microgel
  colloids. \emph{The Journal of chemical physics} \textbf{2004}, \emph{120},
  6197--6206\relax
\mciteBstWouldAddEndPuncttrue
\mciteSetBstMidEndSepPunct{\mcitedefaultmidpunct}
{\mcitedefaultendpunct}{\mcitedefaultseppunct}\relax
\EndOfBibitem
\bibitem[Gnan \latin{et~al.}(2017)Gnan, Rovigatti, Bergman, and
  Zaccarelli]{Gna17}
Gnan,~N.; Rovigatti,~L.; Bergman,~M.; Zaccarelli,~E. In silico synthesis of
  microgel particles. \emph{Macromolecules} \textbf{2017}, \emph{50},
  8777--8786\relax
\mciteBstWouldAddEndPuncttrue
\mciteSetBstMidEndSepPunct{\mcitedefaultmidpunct}
{\mcitedefaultendpunct}{\mcitedefaultseppunct}\relax
\EndOfBibitem
\bibitem[Campbell \latin{et~al.}(2018)Campbell, Saaka, Shao, Gerelli, Cubitt,
  Nazaruk, Matyszewska, and Lawrence]{Ca18}
Campbell,~R.~A.; Saaka,~Y.; Shao,~Y.; Gerelli,~Y.; Cubitt,~R.; Nazaruk,~E.;
  Matyszewska,~D.; Lawrence,~M.~J. Structure of surfactant and phospholipid
  monolayers at the air/water interface modeled from neutron reflectivity data.
  \emph{J. Colloid Interf. Sci.} \textbf{2018}, \emph{531}, 98--108\relax
\mciteBstWouldAddEndPuncttrue
\mciteSetBstMidEndSepPunct{\mcitedefaultmidpunct}
{\mcitedefaultendpunct}{\mcitedefaultseppunct}\relax
\EndOfBibitem
\bibitem[Davies and Rideal(1963)Davies, and Rideal]{Dav63}
Davies,~J.~T.; Rideal,~E.~K. \emph{Interfacial Phenomena}, 2nd ed.; Academic
  Press: San Diego, CA, 1963\relax
\mciteBstWouldAddEndPuncttrue
\mciteSetBstMidEndSepPunct{\mcitedefaultmidpunct}
{\mcitedefaultendpunct}{\mcitedefaultseppunct}\relax
\EndOfBibitem
\bibitem[Cicuta and Terentjev(2005)Cicuta, and Terentjev]{Cicuta2005}
Cicuta,~P.; Terentjev,~E.~M. Viscoelasticity of a protein monolayer from
  anisotropic surface pressure measurements. \emph{The European Physical
  Journal E} \textbf{2005}, \emph{16}, 147–158\relax
\mciteBstWouldAddEndPuncttrue
\mciteSetBstMidEndSepPunct{\mcitedefaultmidpunct}
{\mcitedefaultendpunct}{\mcitedefaultseppunct}\relax
\EndOfBibitem
\bibitem[Campbell \latin{et~al.}(2011)Campbell, Wacklin, Sutton, Cubitt, and
  Fragneto]{Ca11}
Campbell,~R.; Wacklin,~H.; Sutton,~I.; Cubitt,~R.; Fragneto,~G. FIGARO: The new
  horizontal neutron reflectometer at the ILL. \emph{Eur. Phys. J. Plus}
  \textbf{2011}, \emph{126}, 1--22\relax
\mciteBstWouldAddEndPuncttrue
\mciteSetBstMidEndSepPunct{\mcitedefaultmidpunct}
{\mcitedefaultendpunct}{\mcitedefaultseppunct}\relax
\EndOfBibitem
\bibitem[Gutfreund \latin{et~al.}(2018)Gutfreund, Saerbeck, Gonzalez,
  Pellegrini, Laver, Dewhurst, and Cubitt]{Gut18}
Gutfreund,~P.; Saerbeck,~T.; Gonzalez,~M.~A.; Pellegrini,~E.; Laver,~M.;
  Dewhurst,~C.; Cubitt,~R. Towards generalized data reduction on a
  chopper-based time-of-flight neutron reflectometer. \emph{Journal of Applied
  Crystallography} \textbf{2018}, \emph{51}, 606--615\relax
\mciteBstWouldAddEndPuncttrue
\mciteSetBstMidEndSepPunct{\mcitedefaultmidpunct}
{\mcitedefaultendpunct}{\mcitedefaultseppunct}\relax
\EndOfBibitem
\bibitem[Geisel \latin{et~al.}(2012)Geisel, Isa, and Richtering]{Gei12}
Geisel,~K.; Isa,~L.; Richtering,~W. Unraveling the 3D localization and
  deformation of responsive microgels at oil/water interfaces: a step forward
  in understanding soft emulsion stabilizers. \emph{Langmuir} \textbf{2012},
  \emph{28}, 15770--15776\relax
\mciteBstWouldAddEndPuncttrue
\mciteSetBstMidEndSepPunct{\mcitedefaultmidpunct}
{\mcitedefaultendpunct}{\mcitedefaultseppunct}\relax
\EndOfBibitem
\bibitem[Gerelli(2020)]{Ger20}
Gerelli,~Y. Applications of neutron reflectometry in biology. \emph{{EPJ} Web
  of Conferences} \textbf{2020}, \emph{236}, 04002\relax
\mciteBstWouldAddEndPuncttrue
\mciteSetBstMidEndSepPunct{\mcitedefaultmidpunct}
{\mcitedefaultendpunct}{\mcitedefaultseppunct}\relax
\EndOfBibitem
\bibitem[Braslau \latin{et~al.}(1985)Braslau, Deutsch, Pershan, Weiss,
  Als-Nielsen, and Bohr]{Bra85}
Braslau,~A.; Deutsch,~M.; Pershan,~P.~S.; Weiss,~A.~H.; Als-Nielsen,~J.;
  Bohr,~J. Surface Roughness of Water Measured by {X}-Ray Reflectivity.
  \emph{Physical Review Letters} \textbf{1985}, \emph{54}, 114--117\relax
\mciteBstWouldAddEndPuncttrue
\mciteSetBstMidEndSepPunct{\mcitedefaultmidpunct}
{\mcitedefaultendpunct}{\mcitedefaultseppunct}\relax
\EndOfBibitem
\bibitem[Scotti \latin{et~al.}(2020)Scotti, Houston, Brugnoni, Schmidt,
  Schulte, Bochenek, Schweins, Feoktystov, Radulescu, and Richtering]{Sco20ULC}
Scotti,~A.; Houston,~J.; Brugnoni,~M.; Schmidt,~M.; Schulte,~M.; Bochenek,~S.;
  Schweins,~R.; Feoktystov,~A.; Radulescu,~A.; Richtering,~W. Phase behavior of
  ultrasoft spheres show stable bcc lattices. \emph{Physical Review E}
  \textbf{2020}, \emph{102}, 052602\relax
\mciteBstWouldAddEndPuncttrue
\mciteSetBstMidEndSepPunct{\mcitedefaultmidpunct}
{\mcitedefaultendpunct}{\mcitedefaultseppunct}\relax
\EndOfBibitem
\bibitem[Scotti(2021)]{Sco21VF}
Scotti,~A. Characterization of the volume fraction of soft deformable microgels
  by means of small-angle neutron scattering with contrast variation.
  \emph{Soft Matter} \textbf{2021}, \emph{17}, 5548--5559\relax
\mciteBstWouldAddEndPuncttrue
\mciteSetBstMidEndSepPunct{\mcitedefaultmidpunct}
{\mcitedefaultendpunct}{\mcitedefaultseppunct}\relax
\EndOfBibitem
\bibitem[Armanious \latin{et~al.}(2022)Armanious, Gerelli, Micciulla, Pace,
  Welbourn, Sj\"{o}berg, Agnarsson, and H\"{o}\"{o}k]{Arm22}
Armanious,~A.; Gerelli,~Y.; Micciulla,~S.; Pace,~H.~P.; Welbourn,~R. J.~L.;
  Sj\"{o}berg,~M.; Agnarsson,~B.; H\"{o}\"{o}k,~F. Probing the Separation
  Distance between Biological Nanoparticles and Cell Membrane Mimics Using
  Neutron Reflectometry with Sub-Nanometer Accuracy. \emph{Journal of the
  American Chemical Society} \textbf{2022}, \emph{144}, 20726--20738\relax
\mciteBstWouldAddEndPuncttrue
\mciteSetBstMidEndSepPunct{\mcitedefaultmidpunct}
{\mcitedefaultendpunct}{\mcitedefaultseppunct}\relax
\EndOfBibitem
\bibitem[Parratt(1954)]{Par54}
Parratt,~L.~G. Surface Studies of Solids by Total Reflection of X-Rays.
  \emph{Phys. Rev.} \textbf{1954}, \emph{95}, 359--369\relax
\mciteBstWouldAddEndPuncttrue
\mciteSetBstMidEndSepPunct{\mcitedefaultmidpunct}
{\mcitedefaultendpunct}{\mcitedefaultseppunct}\relax
\EndOfBibitem
\bibitem[Abel{\`{e}}s(1950)]{Abe50}
Abel{\`{e}}s,~F. La th{\'{e}}orie g{\'{e}}n{\'{e}}rale des couches minces.
  \emph{Journal de Physique et le Radium} \textbf{1950}, \emph{11},
  307--309\relax
\mciteBstWouldAddEndPuncttrue
\mciteSetBstMidEndSepPunct{\mcitedefaultmidpunct}
{\mcitedefaultendpunct}{\mcitedefaultseppunct}\relax
\EndOfBibitem
\bibitem[Gerelli(2016)]{Ger16}
Gerelli,~Y. Aurore: new software for neutron reflectivity data analysis.
  \emph{J. Appl. Crystallogr.} \textbf{2016}, \emph{49}, 330--339\relax
\mciteBstWouldAddEndPuncttrue
\mciteSetBstMidEndSepPunct{\mcitedefaultmidpunct}
{\mcitedefaultendpunct}{\mcitedefaultseppunct}\relax
\EndOfBibitem
\bibitem[Ninarello \latin{et~al.}(2019)Ninarello, Crassous, Paloli, Camerin,
  Gnan, Rovigatti, Schurtenberger, and Zaccarelli]{Nin19}
Ninarello,~A.; Crassous,~J.~J.; Paloli,~D.; Camerin,~F.; Gnan,~N.;
  Rovigatti,~L.; Schurtenberger,~P.; Zaccarelli,~E. Modeling Microgels with a
  Controlled Structure across the Volume Phase Transition.
  \emph{Macromolecules} \textbf{2019}, \emph{52}, 7584--7592\relax
\mciteBstWouldAddEndPuncttrue
\mciteSetBstMidEndSepPunct{\mcitedefaultmidpunct}
{\mcitedefaultendpunct}{\mcitedefaultseppunct}\relax
\EndOfBibitem
\bibitem[Rovigatti \latin{et~al.}(2015)Rovigatti, {\v{S}}ulc, Reguly, and
  Romano]{rovigatti2015comparison}
Rovigatti,~L.; {\v{S}}ulc,~P.; Reguly,~I.~Z.; Romano,~F. A comparison between
  parallelization approaches in molecular dynamics simulations on GPUs.
  \emph{Journal of computational chemistry} \textbf{2015}, \emph{36},
  1--8\relax
\mciteBstWouldAddEndPuncttrue
\mciteSetBstMidEndSepPunct{\mcitedefaultmidpunct}
{\mcitedefaultendpunct}{\mcitedefaultseppunct}\relax
\EndOfBibitem
\bibitem[Hazra \latin{et~al.}(2023)Hazra, Ninarello, Scotti, Houston,
  Mota-Santiago, Zaccarelli, and Crassous]{Nab23}
Hazra,~N.; Ninarello,~A.; Scotti,~A.; Houston,~J.~E.; Mota-Santiago,~P.;
  Zaccarelli,~E.; Crassous,~J.~J. Structure of Responsive Microgels down to
  Ultralow Cross-Linkings. \emph{Macromolecules} \textbf{2023}, \emph{57},
  339–355\relax
\mciteBstWouldAddEndPuncttrue
\mciteSetBstMidEndSepPunct{\mcitedefaultmidpunct}
{\mcitedefaultendpunct}{\mcitedefaultseppunct}\relax
\EndOfBibitem
\bibitem[Gnan \latin{et~al.}(2017)Gnan, Rovigatti, Bergman, and
  Zaccarelli]{gnan2017silico}
Gnan,~N.; Rovigatti,~L.; Bergman,~M.; Zaccarelli,~E. In silico synthesis of
  microgel particles. \emph{Macromolecules} \textbf{2017}, \emph{50},
  8777--8786\relax
\mciteBstWouldAddEndPuncttrue
\mciteSetBstMidEndSepPunct{\mcitedefaultmidpunct}
{\mcitedefaultendpunct}{\mcitedefaultseppunct}\relax
\EndOfBibitem
\bibitem[Groot and Warren(1997)Groot, and Warren]{groot1997dissipative}
Groot,~R.~D.; Warren,~P.~B. Dissipative particle dynamics: Bridging the gap
  between atomistic and mesoscopic simulation. \emph{The Journal of chemical
  physics} \textbf{1997}, \emph{107}, 4423--4435\relax
\mciteBstWouldAddEndPuncttrue
\mciteSetBstMidEndSepPunct{\mcitedefaultmidpunct}
{\mcitedefaultendpunct}{\mcitedefaultseppunct}\relax
\EndOfBibitem
\bibitem[Camerin \latin{et~al.}(2018)Camerin, Gnan, Rovigatti, and
  Zaccarelli]{camerin2018modelling}
Camerin,~F.; Gnan,~N.; Rovigatti,~L.; Zaccarelli,~E. Modelling realistic
  microgels in an explicit solvent. \emph{Scientific Reports} \textbf{2018},
  \emph{8}, 14426\relax
\mciteBstWouldAddEndPuncttrue
\mciteSetBstMidEndSepPunct{\mcitedefaultmidpunct}
{\mcitedefaultendpunct}{\mcitedefaultseppunct}\relax
\EndOfBibitem
\bibitem[Camerin \latin{et~al.}(2019)Camerin, Fernández-Rodr\'{i}guez,
  Rovigatti, Antonopoulou, Gnan, Ninarello, Isa, and Zaccarelli]{Cam19}
Camerin,~F.; Fernández-Rodr\'{i}guez,~M.~A.; Rovigatti,~L.;
  Antonopoulou,~M.-N.; Gnan,~N.; Ninarello,~A.; Isa,~L.; Zaccarelli,~E.
  Microgels adsorbed at liquid--liquid interfaces: A joint numerical and
  experimental study. \emph{ACS nano} \textbf{2019}, \emph{13},
  4548--4559\relax
\mciteBstWouldAddEndPuncttrue
\mciteSetBstMidEndSepPunct{\mcitedefaultmidpunct}
{\mcitedefaultendpunct}{\mcitedefaultseppunct}\relax
\EndOfBibitem
\bibitem[Camerin \latin{et~al.}(2020)Camerin, Gnan, Ruiz-Franco, Ninarello,
  Rovigatti, and Zaccarelli]{Cam20}
Camerin,~F.; Gnan,~N.; Ruiz-Franco,~J.; Ninarello,~A.; Rovigatti,~L.;
  Zaccarelli,~E. Microgels at interfaces behave as 2D elastic particles
  featuring reentrant dynamics. \emph{Physical Review X} \textbf{2020},
  \emph{10}, 031012\relax
\mciteBstWouldAddEndPuncttrue
\mciteSetBstMidEndSepPunct{\mcitedefaultmidpunct}
{\mcitedefaultendpunct}{\mcitedefaultseppunct}\relax
\EndOfBibitem
\bibitem[Plimpton(1995)]{Pli95}
Plimpton,~S. Fast Parallel Algorithms for Short-Range Molecular Dynamics.
  \emph{Journal of Computational Physics} \textbf{1995}, \emph{117},
  1--19\relax
\mciteBstWouldAddEndPuncttrue
\mciteSetBstMidEndSepPunct{\mcitedefaultmidpunct}
{\mcitedefaultendpunct}{\mcitedefaultseppunct}\relax
\EndOfBibitem
\bibitem[Picard \latin{et~al.}(2017)Picard, Garrigue, Tatry, Lapeyre, Ravaine,
  Schmitt, and Ravaine]{Pic17}
Picard,~C.; Garrigue,~P.; Tatry,~M.-C.; Lapeyre,~V.; Ravaine,~S.; Schmitt,~V.;
  Ravaine,~V. Organization of Microgels at the Air{\textendash}Water Interface
  under Compression: Role of Electrostatics and Cross-Linking Density.
  \emph{Langmuir} \textbf{2017}, \emph{33}, 7968--7981\relax
\mciteBstWouldAddEndPuncttrue
\mciteSetBstMidEndSepPunct{\mcitedefaultmidpunct}
{\mcitedefaultendpunct}{\mcitedefaultseppunct}\relax
\EndOfBibitem
\bibitem[Campbell \latin{et~al.}(2016)Campbell, Tummino, Noskov, and
  Varga]{Cam16}
Campbell,~R.~A.; Tummino,~A.; Noskov,~B.~A.; Varga,~I.
  Polyelectrolyte/surfactant films spread from neutral aggregates. \emph{Soft
  Matter} \textbf{2016}, \emph{12}, 5304--5312\relax
\mciteBstWouldAddEndPuncttrue
\mciteSetBstMidEndSepPunct{\mcitedefaultmidpunct}
{\mcitedefaultendpunct}{\mcitedefaultseppunct}\relax
\EndOfBibitem
\bibitem[Vialetto \latin{et~al.}(2022)Vialetto, Ramakrishna, and Isa]{Via22}
Vialetto,~J.; Ramakrishna,~S.~N.; Isa,~L. In situ imaging of the
  three-dimensional shape of soft responsive particles at fluid interfaces by
  atomic force microscopy. \emph{Science Advances} \textbf{2022}, \emph{8},
  eabq2019\relax
\mciteBstWouldAddEndPuncttrue
\mciteSetBstMidEndSepPunct{\mcitedefaultmidpunct}
{\mcitedefaultendpunct}{\mcitedefaultseppunct}\relax
\EndOfBibitem
\bibitem[Guida \latin{et~al.}(2021)Guida, Aguiar, and Cunha]{Gui21}
Guida,~C.; Aguiar,~A.~C.; Cunha,~R.~L. Green techniques for starch modification
  to stabilize Pickering emulsions: A current review and future perspectives.
  \emph{Current Opinion in Food Science} \textbf{2021}, \emph{38}, 52--61\relax
\mciteBstWouldAddEndPuncttrue
\mciteSetBstMidEndSepPunct{\mcitedefaultmidpunct}
{\mcitedefaultendpunct}{\mcitedefaultseppunct}\relax
\EndOfBibitem
\bibitem[Haupt \latin{et~al.}(2020)Haupt, Medina~Rangel, and Bui]{Hau20}
Haupt,~K.; Medina~Rangel,~P.~X.; Bui,~B. T.~S. Molecularly imprinted polymers:
  Antibody mimics for bioimaging and therapy. \emph{Chemical reviews}
  \textbf{2020}, \emph{120}, 9554--9582\relax
\mciteBstWouldAddEndPuncttrue
\mciteSetBstMidEndSepPunct{\mcitedefaultmidpunct}
{\mcitedefaultendpunct}{\mcitedefaultseppunct}\relax
\EndOfBibitem
\end{mcitethebibliography}

\end{document}